 \documentclass[a4paper]{article}
 \usepackage{etoolbox}
 \usepackage{abstract}
 \newbool{PREPRINT}
 \booltrue{PREPRINT}
\usepackage{makecell}
\usepackage{soul}
\usepackage[colorlinks,bookmarksopen,bookmarksnumbered,citecolor=red,urlcolor=red]{hyperref}
\usepackage{stmaryrd}
\usepackage[usenames]{color}
\usepackage{comment}
\usepackage{amsmath,amssymb}
\usepackage{graphicx}
\usepackage{algorithm}
\usepackage{algorithmicx}
\usepackage{algcompatible}
\usepackage{caption}
\usepackage{listings}
\usepackage{mathtools}
\usepackage{multicol}
\usepackage{placeins}
\usepackage{multirow}
\usepackage{subcaption}
\usepackage{enumitem}
\setlist[description]{itemsep=-0.5ex}
\newcommand{\figdir}{./}

\renewcommand{\vec}[1]{\boldsymbol{#1}}

\newif\ifpreprint
\ifbool{PREPRINT}{ % PREPRINT
\preprinttrue
\usepackage[cm]{fullpage}
\usepackage{amsthm}
\usepackage{multicol}
\setlength\columnsep{20pt}
\usepackage[affil-it]{authblk}

}{ %
\usepackage{fullpage}
\newdefinition{defn}{Definition}
\newdefinition{expl}{Example}
} % PREPRINT

\newcommand{\cellidxi}{\alpha}
\newcommand{\cellidxj}{\beta}

\lstdefinelanguage[ppmd]{python}[]{python}{%
  emph={ParticleLoop,ParticleDat,PositionDat,ScalarArray,GlobalArray,Kernel,PairLoop,Constant,State,Data,IntegratorRange}
}
\definecolor{DarkBlue}{rgb}{0.00,0.00,0.55}
\definecolor{DarkRed}{rgb}{0.55,0.00,0.00}
\definecolor{DarkGreen}{rgb}{0.00,0.55,0.00}
\definecolor{Gray}{rgb}{0.95,0.95,0.95}
\definecolor{Purple}{rgb}{0.5,0.0,0.5}
\definecolor{Bittersweet}{rgb}{1.0,0.44,0.37}
\newcommand{\secapp}{\ifbool{PREPRINT}{Appendix~}{}}
\newcommand{\pparagraph}[1]{\ifbool{PREPRINT}{\paragraph{#1.}}{\paragraph{#1}}}

\title{A new algorithm for electrostatic interactions in Monte Carlo simulations of charged particles}
\ifbool{PREPRINT}{ % PREPRINT
  \author[a]{William~Robert~Saunders}
  \author[b]{James~Grant}
  \author[c,*]{Eike~Hermann~M\"{u}ller}
  \affil[ ]{University of Bath, Bath BA2 7AY, Bath, United Kingdom}
  \affil[a]{Department of Physics}
  \affil[b]{Computing Services}
  \affil[c]{Department of Mathematical Sciences}
\affil[*]{Email: \texttt{e.mueller@bath.ac.uk}}
}{ %
\author[phys]{William~Robert~Saunders}
\ead{w.r.saunders@bath.ac.uk}
\author[comp]{James~Grant}
\ead{r.j.grant@bath.ac.uk}
\author[math]{Eike~Hermann~M\"{u}ller\corref{cor1}\fnref{fn1}}
\ead{e.mueller@bath.ac.uk}
\fntext[fn1]{email: \texttt{e.mueller@bath.ac.uk}}
\cortext[cor1]{Corresponding author}
\address{University of Bath, Bath BA2 7AY, Bath, United Kingdom}
\address[phys]{Department of Physics}
\address[comp]{Computing Services}
\address[math]{Department of Mathematical Sciences}
} % PREPRINT

\begin{document}
\ifbool{PREPRINT}{ % PREPRINT
\maketitle
\begin{onecolabstract}
}{%
\begin{abstract}
} % PREPRINT
To minimise systematic errors in Monte Carlo simulations of charged particles, long range electrostatic interactions have to be calculated accurately and efficiently. Standard approaches, such as Ewald summation or the naive application of the classical Fast Multipole Method, result in a cost per Metropolis-Hastings step which grows in proportion to some positive power of the number of particles $N$ in the system. This prohibitively large cost prevents accurate simulations of systems with a sizeable number of particles. Currently, large systems are often simulated by truncating the Coulomb potential which introduces uncontrollable systematic errors. In this paper we present a new multilevel method which reduces the computational complexity to $\mathcal{O}(\log(N))$ per Metropolis-Hastings step, while maintaining errors which are comparable to direct Ewald summation. We show that compared to related previous work, our approach reduces the overall cost by better balancing time spent in the proposal- and acceptance- stages of each Metropolis-Hastings step. By simulating large systems with up to $N=10^5$ particles we demonstrate that our implementation is competitive with state-of-the-art MC packages and allows the simulation of very large systems of charged particles with accurate electrostatics.
\ifbool{PREPRINT}{ % PREPRINT
\end{onecolabstract}
% === ACM classifiers ===
% J.2 PHYSICAL SCIENCES AND ENGINEERING -> Chemistry, Physics
% F.2.1 ANALYSIS OF ALGORITHMS AND PROBLEM COMPLEXITY -> Numerical Algorithms and Problems
% G.1: NUMERICAL ANALYSIS
% === MSC classifiers ===
% 78M16 Optics, Electromagnetic Theory -> Multipole methods
% 82C80 STATISTICAL MECHANICS, STRUCTURE OF MATTER -> Numerical methods (Monte Carlo, series, etc.)
% 65Y20 NUMERICAL ANALYSIS -> Complexity and performance of numerical algorithms

\textbf{keywords}:
\newcommand{\sep}{, }
}{
\end{abstract}
\begin{keyword}
} % PREPRINT
Monte Carlo\sep electrostatics\sep particle simulations\sep computational complexity\sep Fast Multipole Method
\ifbool{PREPRINT}{ % PREPRINT
\\[1ex]
}{
\end{keyword}
\maketitle
}
%%%%%%%%%%%%%%%%%%%%%%%%%%%%%%%%%%%%%%%%%%%%%%%%%%%%%%%%%%%%%%%%%%%
\section{Introduction}
%%%%%%%%%%%%%%%%%%%%%%%%%%%%%%%%%%%%%%%%%%%%%%%%%%%%%%%%%%%%%%%%%%%
The accurate representation of all pairwise interactions in classical atomistic simulations is important to minimise systematic errors. In this paper we focus on Monte Carlo (MC) simulations of charged particles. Short-range interactions such as the Lennard-Jones potential can be safely truncated at a finite cutoff distance: when calculating energy differences in a proposed MC move only interactions with a fixed number of other particles in close proximity of the moving particle need to be taken into account. For fixed density the cost of one local Metropolis-Hastings (MH) step is constant, independent of the total number $N$ of particles in the system. However, due to the long-range nature of the Coulomb potential, which decays in proportion to the inverse distance between two charges, including electrostatics is far from trivial since interactions with \textit{all} other particles in the system have to be considered. Worse, interactions with periodic copies or mirror charges have to be taken into account if non-trivial boundary conditions are used.

As the review in \cite{Cisneros2013} shows, a plethora of methods have been developed to address this issue, but care has to be taken to obtain reliable results. In this context the authors of \cite{Jorge2002} for example find that truncating the Coulomb potential in the MC simulation of water in a highly anisotropic geometry leads to significant systematic errors. Other methods which have been proposed to avoid this problem include solving the Poisson equation with a grid-based multigrid method \cite{Sandalci1997}, Ewald summation \cite{Ewald1921} or the naive application of the Fast Multipole Method (FMM) \cite{Greengard1987,Greengard1988,Greengard1997}. Unfortunately all those approaches result in a prohibitively large computational cost as the number of particles $N$ in the system grows, typically the cost per MH step increases as $\mathcal{O}(N)$ or $\mathcal{O}(\sqrt{N})$. This renders the simulation of very large systems impossible and limits the predictive power of computer experiments.

This is particularly keenly felt in MC simulations, where in contrast to molecular dynamics, systems evolve through discontinuous moves of individual atoms or small numbers of particles.  The current inability to treat electrostatics accurately and efficiently prohibits the application of MC to problems with larger particle counts. On the other hand, in many circumstances it is desirable to simulate a system with MC instead of molecular dynamics.  In particular, the grand canonical or semi-grand ensembles where the number of particles can fluctuate are only accessible with MC \cite{Crabtree2013,Pham2020}. Exchanging particle position in so-called `swap moves' can be far more efficient at equilibrating systems e.g. when simulating the distribution of impurities near grain boundaries.  The lack of efficient electrostatic algorithms also limits the use of methods such as phase- and lattice switch MC \cite{Bruce2000,Underwood2015} which allow the accurate determination of free energy differences between phases.  Enabling the routine application of MC to larger systems of charged particles will allow researchers to study the physics of a particular system with the computationally most appropriate evolution algorithm.

In this paper we show how the limitations of the MC method for charged particle systems can be overcome by constructing an algorithm which reduces the cost per MH step to $\mathcal{O}(\log(N))$ without sacrificing accuracy, thereby making much larger simulations with realistic electrostatics feasible. Our multilevel approach is inspired by the Fast Multipole method and similar to the method recently proposed in \cite{Hoeft2017}. However, compared to \cite{Hoeft2017} it leads to an overall reduction of computational cost by balancing the time spent in proposing and accepting MC moves in realistic MC simulations.

The key observation motivating our new method is the following: standard FMM constructs local expansions for evaluating the long range interactions on the finest level of the hierarchical tree. While the evaluation of those expansions (and direct interactions with all close by neighbours) in the proposal stage of a MH step is $\mathcal{O}(1)$, re-calculating the local expansions incurs a cost of $\mathcal{O}(N)$ since all local expansions are re-calculated in the upward and downward pass of the algorithm, resulting in an overall $\mathcal{O}(N)$ cost per MH step. By storing the local expansions on each level of the multilevel hierarchy instead of accumulating them on the finest level, the relative cost of the proposal- and accept- stage can be balanced since each of those two steps requires a fixed number of operations on each level of the multilevel hierarchy. As the number of levels $L$ is proportional to $\log(N)$, this results in a total computational complexity per MH step which grows logarithmically in the number of particles.

In summary, the new contributions of our work are as follows:
\begin{enumerate}
\item We describe a new hierarchical algorithm for Monte Carlo simulations with accurate electrostatics, which has an $\mathcal{O}(\log(N))$ computational complexity per Metropolis-Hastings step.
\item By comparing to the similar method in \cite{Hoeft2017} we show that our algorithm leads to an overall reduction in cost for realistic Monte Carlo simulations.
  \item We describe the efficient implementation of our algorithm in the performance-portable PPMD framework \cite{Saunders2018} recently developed in our group. Since it has a user-friendly high-level Python interface, PPMD allows the easy implementation of Monte Carlo algorithms with accurate electrostatics.
\item We demonstrate that the runtime of our implementation is competitive with the state-of-the-art MC code DL\_MONTE \cite{Purton2013,Brukhno2019}, which struggles to simulate systems of the size that are easily accessible with our implementation. 
\end{enumerate}

For systems with $N=10^5$ particles our implementation is an order of magnitude faster than the DL\_MONTE code.
At this system size, we observe that the alternative approach in \cite{Hoeft2017} results in a 30\% longer simulation time than our method. By fitting a semi-empirical model to predict the cost of a simulation as a function of the problem size, we show that asymptotically (i.e. for $N\rightarrow\infty$) we expect our algorithm to be twice as fast as the one in \cite{Hoeft2017}.
\pparagraph{Structure} This paper is organised as follows: After discussing related work in Section \ref{sec:relatedwork}, we review the native FMM algorithm and describe our new method in Section \ref{sec:method}, where we also compare it to the approach in \cite{Hoeft2017}. Following a description of the Python interface for our implementation of the algorithms introduced in this paper in Section \ref{sec:implementation}, numerical results are presented in Section \ref{sec:results}. We conclude and outline directions for further work in Section \ref{sec:conclusion}.
%%%%%%%%%%%%%%%%%%%%%%%%%%%%%%%%%%%%%%%%%%%%%%%%%%%%%%%%%%%%%%%%%%%
\section{Related work}\label{sec:relatedwork}
%%%%%%%%%%%%%%%%%%%%%%%%%%%%%%%%%%%%%%%%%%%%%%%%%%%%%%%%%%%%%%%%%%%
Methods for including untruncated electrostatic interactions in MC simulations with a computational complexity of $\mathcal{O}(\log(N))$ per MH step have been developed previously in \cite{Hoeft2017,Gan2014}. Compared to our approach, the method in \cite{Hoeft2017} does not construct local expansions, thereby avoiding their recalculation whenever a particle is moved. While this might look like a reasonable simplification, it actually makes evaluating the change in potential for each proposed (but potentially rejected) MC transition more expensive. Since there is typically more than one proposed move per accepted transition, this renders the method in \cite{Hoeft2017} more expensive overall, as our numerical experiments confirm.

A modification of the Barnes-Hut octree algorithm is discussed in \cite{Gan2014}. Similar to FMM, the classical Barnes-Hut method constructs a hierarchical mesh structure, and represents the distribution of particles in cells on each level by their multidimensional Taylor expansion coefficients. While the calculation of the total electrostatic energy with the octree algorithm is $\mathcal{O}(N\log(N))$, the authors of \cite{Gan2014} present a modification of the method which has a cost of $\mathcal{O}(\log(N))$ per local MC step.

The $\mathcal{O}(\log(N))$ algorithms presented here and in \cite{Hoeft2017,Gan2014} improve on what can be achieved with Ewald summation \cite{Ewald1921,Frenkel2001}, for which the change in electrostatic energy per MC proposal can be calculated at a computational complexity of $\mathcal{O}(\sqrt{N})$. This is because the overall $\mathcal{O}(N^{3/2})$ cost of the Ewald-based energy calculation is made up by an iteration over all $N$ particles and a sum over $O(\sqrt{N})$ reciprocal vectors (long-range contribution) and neighbouring particles (short-range contribution). If only $\mathcal{O}(1)$ particles move in each proposed move, only a small number of the $\mathcal{O}(\sqrt{N})$ sums have to be evaluated. A similar approach is currently explored in the DL\_MONTE code \cite{Purton2013,Brukhno2019}, though the implementation at present is limited by the fact that the short-range cutoff of the Ewald summation has to be identical to the cutoff of all other local interactions. In this paper we present numerical results which show that our new method can be used to simulate systems with up to $10^5$ charges and accurate electrostatic interactions at a cost of around $1\text{ms}$ per MH step.

For completeness, it should be pointed out that other methods with a computational complexity of $\mathcal{O}(N)$ per MH step have also been developed. For example, in \cite{Sandalci1997} a multigrid method is used to solve the Poisson equation to obtain the electrostatic potential generated by the particles. Since the global electrostatic field has to be re-computed for each (local) particle move and the cost of multigrid grows in proportion to the number of grid points \cite{Trottenberg2000}, which in turn has to increase with the problem size to ensure an accurate representation of the charge distribution on the computational mesh, the cost is inherently $\mathcal{O}(N)$. An additional disadvantage of the method is that the mapping of the highly peaked charge distribution to a grid introduces interpolation errors.

While all methods discussed so far focus on the development of fast methods for computing the electrostatic interactions between particles based on Gauss' law and the fact that the electric field is the gradient of a scalar potential, a radically different approach is pursued in \cite{Rottler2004}. The authors of this paper argue that (statistically) integrating over a ficticious traverse electric field in addition to the different particle configurations produces the same results as a traditional Monte Carlo simulation with standard Coulombic interactions. While this enlarges the configuration space, it crucially only requires local updates for each particle move. Since the traverse electric field relaxes rapidly, this only introduces a small overhead, independent of the problem size. Overall the cost of updating all particle positions and the traverse field grows in proportion to $N$, which results in a $\mathcal{O}(1)$ cost per individual particle move. A non-trivial interpolation scheme is used in \cite{Rottler2004} to reduce lattice artifacts that are introduced by enforcing Gauss' law on a Cartesian grid.
%%%%%%%%%%%%%%%%%%%%%%%%%%%%%%%%%%%%%%%%%%%%%%%%%%%%%%%%%%%%%%%%%%%
\section{Method}\label{sec:method}
%%%%%%%%%%%%%%%%%%%%%%%%%%%%%%%%%%%%%%%%%%%%%%%%%%%%%%%%%%%%%%%%%%%
We now discuss the new approach introduced in this work. After briefly reviewing key concepts of the classical FMM algorithm we describe our method in detail in Section \ref{sec:MC_FMM_method} and compare it to the related technique in \cite{Hoeft2017} in Section \ref{sec:alternative_multipole_algorithm}.
%%%%%%%%%%%%%%%%%%%%%%%%%%%%%%%%%%%%%%%%%%%%%%%%%%%%%%%%%%%%%%%%%%%
\subsection{Fast Multipole method}
\begin{figure}
  \begin{center}
    \includegraphics[width=0.6\linewidth]{\figdir/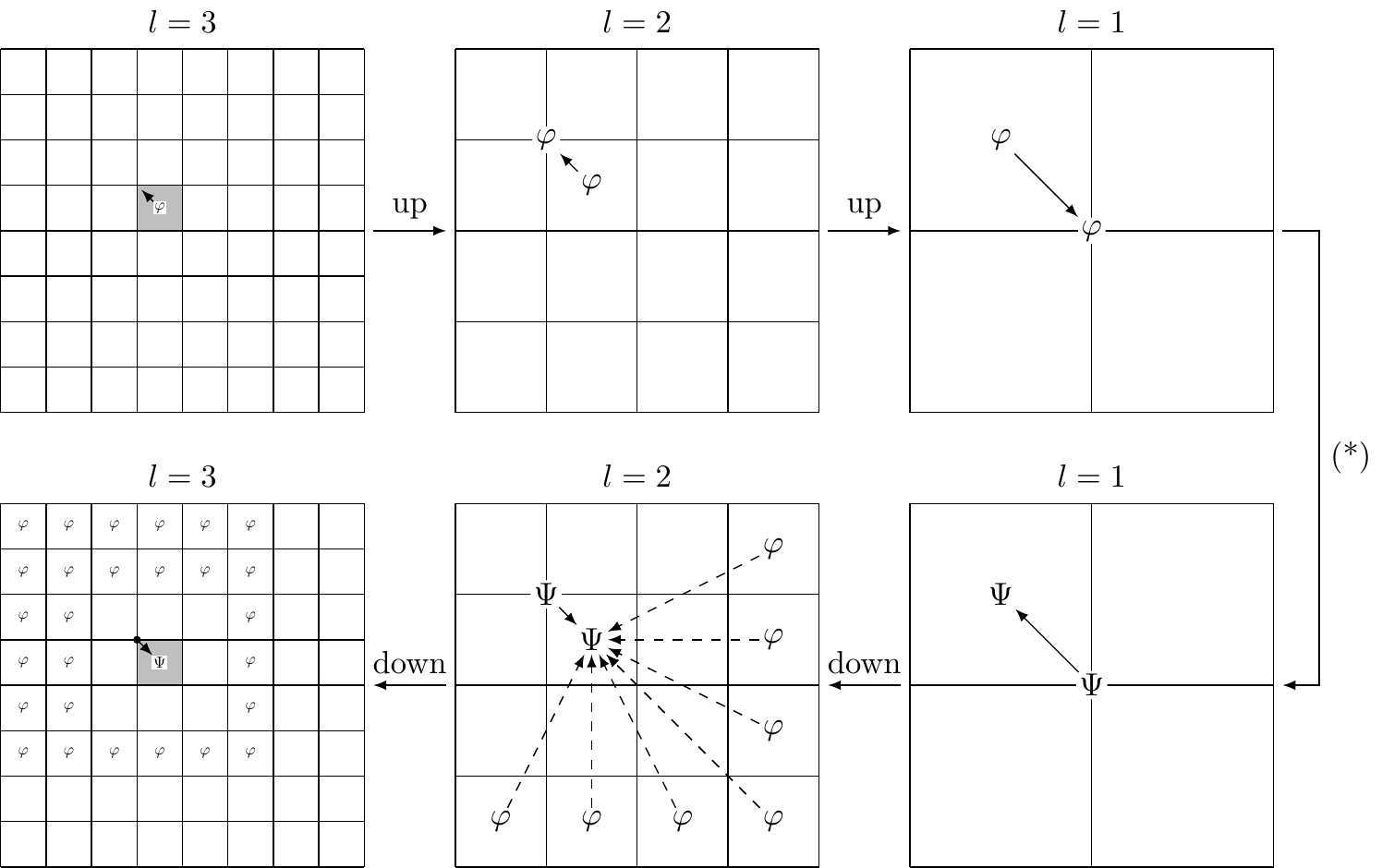}
    \caption{Schematic illustration of the classical Fast Multipole Method in two dimensions first described in \cite{Greengard1987,Greengard1988}. The number of levels is $L=4$ in this example. The upward pass for constructing the multipole expansions ($\varphi$) is shown in the top row, while the local expansions ($\Psi$) are built in the downward pass at the bottom of the figure. The asterisk (*) on the right hand side denotes special operations on the coarsest level for incorporating (potentially nontrivial) boundary conditions.}
    \label{fig:FMMschematic}
  \end{center}
\end{figure}

%%%%%%%%%%%%%%%%%%%%%%%%%%%%%%%%%%%%%%%%%%%%%%%%%%%%%%%%%%%%%%%%%%%
In three dimensions the FMM algorithm \cite{Greengard1987,Greengard1988,Greengard1997} uses a hierarchical grid with $L$ levels for the computational domain $\Omega$ (which is assumed to be a cube of width $a$) such that the number of cells on each level $\ell$ is $M_\ell = 8^{\ell-1}$ for $\ell=1,\dots,L$. The number of cells on the finest level is $M=M_L$, and (to balance the work between the long range and direct field calculation) typically $L$ is chosen such that there are $\mathcal{O}(1)$ particles in each fine level cell. Each cell on level $\ell=1,\dots,L-1$ is subdivided into 8 child-cells on the next-finer level; conversely each cell on level $\ell=2,\dots,L$ has a unique parent cell. The Fast Multipole Algorithm now computes the electrostatic potential by splitting it into two contributions. First, the long range part is calculated by working out the multipole expansion of all charges in a fine level cell and transforming them into multipole expansions on the coarser levels in the upward pass of the algorithm. In the downward pass the multipole expansions on each level are transformed into local expansions around the centre of a cell. Those are then recursively combined into local expansions in the child cells. By only considering the contribution from multipole expansions in a fixed number of well-separated cells on each level, the potential of distant charges is resolved at the appropriate level of accuracy, while including the contribution from closer charges on finer levels. The method for calculating the long range contribution is shown schematically in Figure \ref{fig:FMMschematic}; we refer the reader to \cite{Greengard1987,Greengard1988,Greengard1997} for further details.
For the following discussion of our FMM variant for MC simulations the notion of an \textit{interaction list} ($\operatorname{IL}$) of a particular cell is crucial. For a cell $\cellidxi$ on level $\ell$ this interaction list $\text{IL}(\cellidxi)$ is the set of cells which are the children of the parent cell of $\cellidxi$ and its nearest neighbours, but which are well separated from $\cellidxi$, i.e. not direct neighbours of $\cellidxi$ on level $\ell$. Explicitly, the interaction list is defined as
\begin{equation*}
\operatorname{IL}(\cellidxi) = \textsf{children}\left(\mathcal{N}_b\left(\textsf{parent}(\cellidxi)\right)\right)\backslash\left(\cellidxi\cup\mathcal{N}_b(\cellidxi)\right),
\end{equation*}
where $\mathcal{N}_b(\cellidxi)$ is the set of the 26 nearest neighbours of the cell $\cellidxi$; the functions $\textsf{children}(\cellidxi)$, $\textsf{parent}(\cellidxi)$ return the set of child cells or the parent cell of a particular cell $\cellidxi$. An example of an interaction list can be found in the bottom left corner of Figure \ref{fig:FMMschematic}: all cells labelled with the letter $\varphi$ are in the interaction list of the gray cell labelled with a $\Psi$.

Finally, interactions with charges in neighbouring fine level cells are included by directly evaluating the $1/r$ potential generated by those charges.
%%%%%%%%%%%%%%%%%%%%%%%%%%%%%%%%%%%%%%%%%%%%%%%%%%%%%%%%%%%%%%%%%%%
\subsection{FMM for Monte Carlo simulations}\label{sec:MC_FMM_method}
\begin{figure}
  \begin{center}
    \includegraphics[width=0.95\linewidth]{\figdir/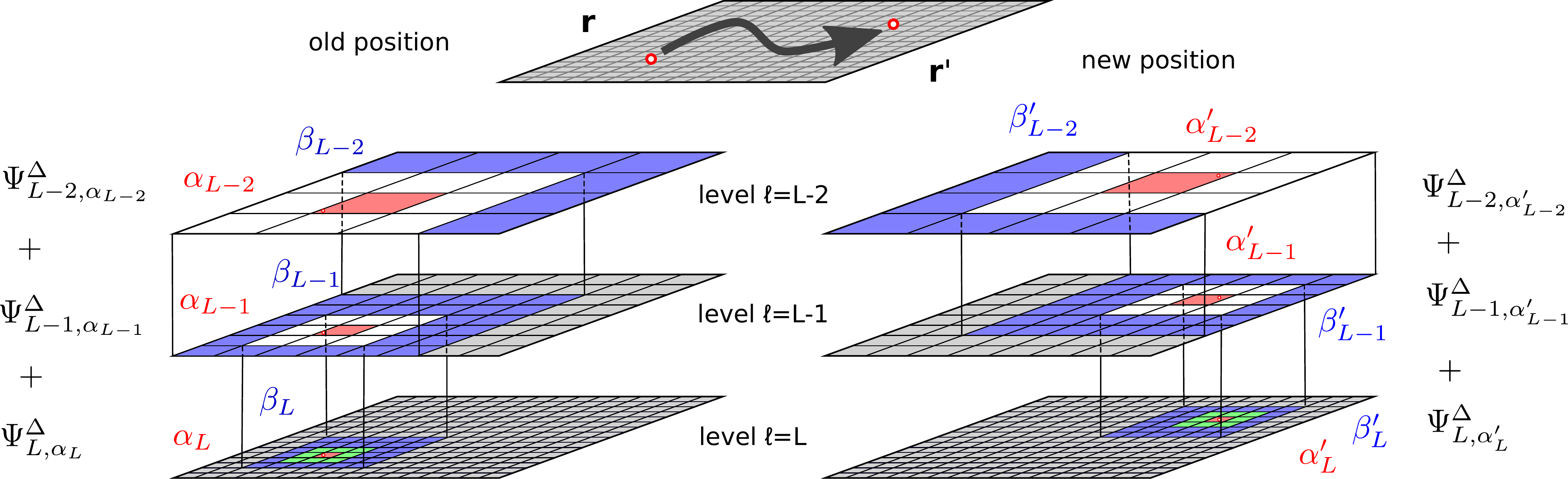}
    \caption{Schematic illustration of the new method for a proposing a MC move $\vec{r}\rightarrow\vec{r}'$, as described in Section \ref{sec:MC_FMM_method}. On each level $\ell$ the cell $\cellidxi_\ell$ containing $\vec{r}$ and the corresponding cell $\cellidxi_\ell$ containing $\vec{r}'$ are marked in red. Cells $\cellidxj_\ell$ in the interaction list of $\cellidxi_\ell$ and cells $\cellidxj'_\ell$ in the interaction list of $\cellidxi_\ell$ are shown in blue. Interactions with particles in green cells have to be evaluated directly when a move is proposed.}
    \label{fig:method_schematic}
  \end{center}
\end{figure}

%%%%%%%%%%%%%%%%%%%%%%%%%%%%%%%%%%%%%%%%%%%%%%%%%%%%%%%%%%%%%%%%%%%
Now consider the following modification of FMM. Let $\Psi^{\Delta}_{\ell,\cellidxi}$ be the $p$-term local expansion around the center of cell $\cellidxi$ on level $\ell$ such that $\Psi^\Delta_{\ell,\cellidxi}$ contains contributions from all charges in the interaction list $\operatorname{IL}(\cellidxi)$ of $\cellidxi$. Note that this is different from standard FMM, where the local expansions $\Psi_{\ell,\cellidxi}$ contain the contribution of all charges not contained in the cell $\cellidxi$ or its 26 nearest neighbours. However, $\Psi_{\ell,\cellidxi}$ can be obtained by summing the $\Psi^\Delta_{\ell,\cellidxi}$ on the current and coarser levels, namely
\begin{equation}
  \Psi_{\ell,\cellidxi} = \sum_{\ell'=1}^{\ell} \Psi^{\Delta}_{\ell',\cellidxi_{\ell'}} \qquad\text{with $\cellidxi_\ell = \cellidxi$ and $\cellidxi_{\ell'}=\textsf{parent}(\cellidxi_{\ell'+1})$ for all $\ell'=1,\dots,\ell-1$}.
\label{eqn:DeltaPsiSum}
\end{equation}
For a cell $\cellidxi$ on level $\ell$ the local expansion $\Psi^\Delta_{\ell,\cellidxi}$ can be expressed in terms of the coefficients $(L^\Delta_{\ell,\cellidxi})_n^m$ as 
\begin{equation}
  \Psi^\Delta_{\ell,\cellidxi}(\vec{\delta r}) = \sum_{n=0}^{p}\sum_{m=-n}^{+n} \left(L^\Delta_{\ell,\cellidxi}\right)_n^m (\delta r)^n Y_n^m(\delta \theta,\delta\phi)
  \qquad\text{with $(\delta r,\delta\theta,\delta\phi)=\textsf{spherical}(\vec{\delta r})$}.
  \label{eqn:local_expansion}
\end{equation}
Here $\vec{\delta r}$ is the position of the particle measured relative to the centre $\vec{R}_{\cellidxi}$ of the cell $\cellidxi$. The function $\textsf{spherical}(\vec{r})$ returns the spherical coordinates $(r,\theta,\phi)$ of a vector $\vec{r}$.
We further call the set $\textsf{ancestors}(\cellidxi)=\{\cellidxi_{\ell-1},\cellidxi_{\ell-2},\dots,\cellidxi_2,\cellidxi_1\}$ defined recursively in Equation \eqref{eqn:DeltaPsiSum} the \textit{ancestors} of cell $\cellidxi$.
Our strategy for evaluating the long range contributions in Monte Carlo simulations is as follows (see Figure \ref{fig:method_schematic}):
\begin{description}
  \item[Initialisation.] At the beginning of the simulation, calculate the local expansion coefficients $(L^\Delta_{\ell,\cellidxi})_n^m$ for all cells $\cellidxi$ and on all levels $\ell$ by using a slightly modified variant of the upward/downward pass in the Fast Multipole Algorithm.
  \item[Proposal.] Consider a proposed MC move $\vec{r}\rightarrow\vec{r}'$ of charge $q$ such that the original position $\vec{r}$ is contained in the fine-level cell $\cellidxi$ and the new position $\vec{r}'$ in the fine-level cell $\cellidxi'$ (which could be identical to $\cellidxi$). To evaluate the change in the long range potential, evaluate and accumulate the $\Psi^\Delta_{\ell,\cellidxi_\ell}(\vec{r}-\vec{R}_{\cellidxi_\ell})$ and $\Psi^\Delta_{\ell,\cellidxi'_\ell}(\vec{r}'-\vec{R}_{\cellidxi_\ell'})$ on all levels $\ell=1,\dots,L$ of the hierarchy by using the sum in Equation \eqref{eqn:DeltaPsiSum}, where $\vec{R}_\alpha$ is the centre of cell $\alpha$. This gives the change in long-range electrostatic energy $\Delta U_{\mathrm{lr}}=q(\Psi_{L,\cellidxi'}(\vec{r}'-\vec{R}_{\cellidxi'})-\Psi_{L,\cellidxi}(\vec{r}-\vec{R}_{\cellidxi}))$. Secondly, add the change in short-range energy $\Delta U_{\mathrm{direct}}$ from a direct calculation of the interactions with particles in the same and adjacent cells. Finally, remove the spurious self-interaction contribution $q^2 / |\vec{r}' - \vec{r}|$ which is due to the potential field induced by the charge at the original position. This self-interaction correction is described in detail in \cite[Section 3]{Saunders2019}.
  \item[Accept.] Assume we accept a move $\vec{r}\rightarrow\vec{r}'$ of charge $q$ such that $\vec{r}$ lies in cell $\cellidxi$ and $\vec{r}'$ in cell $\cellidxi'$. For all cells $\cellidxj$ in the interaction list of $\cellidxi$ and all its ancestors the contribution of a monopole with charge $q$ is subtracted from $\Psi^\Delta_{\ell,\cellidxj_{\ell}}$ on all levels $\ell=1,\dots,L$. This requires updating the expansion coefficients $L^\Delta_{\ell,\cellidxi_{\ell}}$. Conversely, a monopole of charge $q$ is added to the local expansions $\Psi^\Delta_{\ell,\cellidxj'_{\ell}}$ of for all cells $\cellidxj'_\ell$ in the interaction list of $\cellidxi'$ and its ancestors.
\end{description}
The \textit{propose} and \textit{accept} steps are written down explicitly in Algorithms \ref{alg:propose_local} and \ref{alg:accept_local}; the direct calculation of the interaction with particles in the same fine-level cell or directly adjacent cells to obtain $\Delta U_{\mathrm{direct}}$ is given in Algorithm \ref{alg:propose_direct}. In Algorithm \ref{alg:propose_direct} we remove the spurious self-interaction that occurs between the charge at the proposed position with itself at the original position.
\renewcommand\thealgorithm{L.\arabic{algorithm}}
\begin{algorithm}
  \caption{Propose a move $\vec{r}\rightarrow\vec{r}'$ using local expansions. Input: initial position $\vec{r}$ of particle with charge $q$; target position $\vec{r}'$. Output: change in total electrostatic energy $\Delta U$}
  \label{alg:propose_local}
  \begin{algorithmic}[1]
    \STATE{Find fine-level cells $\cellidxi_L, \cellidxi'_L$ such that $\vec{r}\in\cellidxi_L$ and $\vec{r}'\in\cellidxi'_L$}
    \STATE{$\Delta U\mapsfrom 0$}
    \FOR{$\ell=L,L-1,\dots,1$}
    \STATE{Update $\Delta U\mapsfrom\Delta U + q\left(\Psi^{\Delta}_{\ell,\cellidxi'_{\ell}}(\vec{r}'-\vec{R}_{\cellidxi'_\ell})- \Psi^{\Delta}_{\ell,\cellidxi_{\ell}}(\vec{r}-\vec{R}_{\cellidxi_\ell})\right)$ using \textit{local} expansion in Equation \eqref{eqn:local_expansion}}
    \IF{$\ell>1$}
    \STATE{Set $\cellidxi_{\ell-1}\mapsfrom \textsf{parent}(\cellidxi_\ell)$; $\cellidxi'_{\ell-1}\mapsfrom \textsf{parent}(\cellidxi'_\ell)$}
    \ENDIF
    \ENDFOR
    \STATE{Calculate change $\Delta U_{\text{direct}}$ in electrostatic energy from direct interactions with Algorithm \ref{alg:propose_direct}}
    \STATE{$\Delta U\mapsfrom\Delta U+\Delta U_{\text{direct}}$}
  \end{algorithmic}
\end{algorithm}
\begin{algorithm}
  \caption{Accept a move $\vec{r}\rightarrow\vec{r}'$ using local expansions. Input: old position $\vec{r}$, new position $\vec{r}'$}
  \label{alg:accept_local}
  \begin{algorithmic}[1]
    \STATE{Find fine-level cells $\cellidxi_L, \cellidxi'_L$ such that $\vec{r}\in\cellidxi_L$ and $\vec{r}'\in\cellidxi'_L$}
    \FOR{$\ell=L,L-1,\dots,1$}
    \FOR{$\cellidxj_\ell\in\text{IL}(\cellidxi_\ell)$ and $\cellidxj'_\ell\in\text{IL}(\cellidxi'_\ell)$}
    \STATE{Set $(\delta r,\delta\theta,\delta\phi) \mapsfrom \textsf{spherical}(\vec{r}-\vec{R}_{\cellidxj_\ell})$ and $(\delta r',\delta\theta',\delta\phi')\mapsfrom\textsf{spherical}(\vec{r}'-\vec{R}_{\cellidxj'_\ell})$}
    \FOR{$n=0,\dots,p$}
    \FOR{$m=-n,\dots,+n$}
    \STATE{Update $(L^\Delta_{\ell,\cellidxj_\ell})_n^m \mapsfrom (L^\Delta_{\ell,\cellidxj_\ell})_n^m - q (\delta r)^{-(n+1)} Y_n^{-m}(\delta\theta,\delta\phi)$}
    \STATE{Update $(L^\Delta_{\ell,\cellidxj'_\ell})_n^m \mapsfrom (L^\Delta_{\ell,\cellidxj'_\ell})_n^m + q (\delta r')^{-(n+1)}Y_n^{-m}(\delta\theta',\delta\phi')$}
    \ENDFOR
    \ENDFOR
    \ENDFOR
    \IF{$\ell>1$}
    \STATE{Set $\cellidxi_{\ell-1}\mapsfrom \textsf{parent}(\cellidxi_\ell)$; $\cellidxi'_{\ell-1}\mapsfrom \textsf{parent}(\cellidxi'_\ell)$}
    \ENDIF
    \ENDFOR
  \end{algorithmic}
\end{algorithm}
\begin{algorithm}
  \caption{Initialise local expansion coefficients $(L^\Delta_{\ell,\cellidxi})_n^m$ for electrostatic calculation with Algorithms \ref{alg:propose_local} and \ref{alg:accept_local}}
  \label{alg:initialise_local}
  \begin{algorithmic}[1]
  \FOR{levels $\ell=1,\dots,L$}
  \FORALL{cells $\cellidxi_\ell$ on level $\ell$}
  \FOR{$n=0,\dots,p$ and $m=-n\dots+n$}
  \STATE{Set $(L^\Delta_{\ell,\cellidxi_\ell})_n^m = 0$}
  \ENDFOR
  \ENDFOR
  \FORALL{cells $\cellidxi_\ell$ on level $\ell$}
  \FORALL{particles with charge $q_i$ and position $\vec{r}_i\in\cellidxi_\ell$}
  \FORALL{cells $\cellidxj_\ell\in\text{IL}(\cellidxi_\ell)$}
  \STATE{Set $(\delta r_i,\delta\theta_i,\delta\phi_i)\mapsfrom\textsf{spherical}(\vec{r}_i-\vec{R}_{\cellidxj_\ell})$}
  \FOR{$n=0,\dots,p$}
  \FOR{$m=-n,\dots,+n$}
  \STATE{Update $(L^\Delta_{\ell,\cellidxj_\ell})_n^m \mapsfrom (L^\Delta_{\ell,\cellidxj_\ell})_n^m + q_i (\delta r_i)^{-(n+1)}Y_n^{-m}(\delta\theta_i,\delta\phi_i)$}
  \ENDFOR
  \ENDFOR
  \ENDFOR
  \ENDFOR
  \ENDFOR
  \ENDFOR
  \end{algorithmic}
\end{algorithm}
\renewcommand{\thealgorithm}{\arabic{algorithm}}
\begin{algorithm}
  \caption{Calculate change in electrostatic energy from direct interactions for a proposed move $\vec{r}\rightarrow\vec{r}'$. Input: initial position $\vec{r}\in\cellidxi_L$ of particle with charge $q$; target position $\vec{r}'\in\cellidxi'_L$. Output: change in direct electrostatic interaction energy $\Delta U_{\text{direct}}$}
  \label{alg:propose_direct}
  \begin{algorithmic}[1]
    \STATE{$\Delta U_{\text{direct}}\mapsfrom 0$}
    \FORALL{particles with charge $q_i$ and position $\vec{r}_i\in\cellidxi_L\cup \mathcal{N}_b(\cellidxi_L)$, $\vec{r}_i\ne\vec{r}$}
      \STATE{Update $\Delta U_{\text{direct}} \mapsfrom \Delta U_{\text{direct}} - \frac{qq_i}{|\vec{r}-\vec{r}_i|}$}
    \ENDFOR
    \FORALL{particles with charge $q'_i$ and position $\vec{r}'_i\in\cellidxi'_L\cup \mathcal{N}_b(\cellidxi'_L)$, $\vec{r}_i'\ne\vec{r}'$}
      \STATE{Update $\Delta U_{\text{direct}} \mapsfrom \Delta U_{\text{direct}} + \frac{qq'_i}{|\vec{r}'-\vec{r}'_i|}$}
    \ENDFOR
      \STATE{Remove self-interaction  $\Delta U_{\text{direct}} \mapsfrom \Delta U_{\text{direct}} - \frac{q^2}{|\vec{r}' - \vec{r}|}$}
  \end{algorithmic}
\end{algorithm}

Since the local expansions with $\mathcal{O}(p^2)$ terms need to be evaluated in two cells per level, the cost of one proposal is $\mathcal{O}(p^2L)=\mathcal{O}(p^2\log(N))$. Similarly, when updating the $\mathcal{O}(p^2)$ expansion coefficients $L^\Delta_{\ell,\cellidxi_\ell}$ while accepting a move, the number of cells in the interaction list on each level is constant ($6^d-3^d=189$ in $d=3$ dimensions, to be specific). Therefore the computational complexity of the accept stage is also $\mathcal{O}(p^2L)=\mathcal{O}(p^2\log(N))$. The larger constant (compared to the one in the propose stage) arises due to the fact that $2\times189=378$ instead of $2$ cells have to be considered on each level in this stage and is partially compensated by two effects:
\begin{enumerate}
\item Typically only a fraction of all proposed moves are accepted. 
\item Each MC proposal also requires the calculation of the short-range electrostatic interactions, which is not necessary in the accept stage.
\end{enumerate}
The short-range contribution of the electrostatic potential is evaluated by calculating the contribution of all charges in $\mathcal{N}_b(\cellidxi)$ and $\mathcal{N}_b(\cellidxi')$ directly in Algorithm \ref{alg:propose_direct}. As in the standard FMM algorithm the number of charges per fine level cell is constant and independent of the number of levels; each cell typically contains at the order of 1-10 charges. This guarantees that the total cost of the electrostatic calculation in the proposal step is still $\mathcal{O}(\log(N))$ after the direct, short-range interactions are included using Algorithm \ref{alg:propose_direct}.

We conclude that the computational complexity of the electrostatics in one MH step, which consists of a proposed move, potentially followed by one accepted transition, is $\mathcal{O}(p^2\log(N))$.

The initialisation of the local expansions $\Psi^\Delta_{\ell,\cellidxi}$ at the beginning of the simulation could be carried out in $\mathcal{O}(p^4N)$ time with a minimally modified variant of the standard FMM algorithm, which is written down explicitly as Algorithm 2 in \cite{Saunders2019}. Apart from renaming $L_{\ell,\cellidxi}\mapsto L^{\Delta}_{\ell,\cellidxi}$ in the local expansions, the only difference is that line 17 of this algorithm has to be replaced by $\Psi_{\ell,\cellidxi}\mapsfrom 0$ and the loop over cells to construct $\overline{\Psi}_{\ell,\cellidxi}$ in lines 13-15 is no longer necessary. However, since the setup cost is amortised anyway by the large number of MH steps, we chose a slightly more expensive but simpler approach, which is written down in Algorithm \ref{alg:initialise_local} and avoids the calculation of multipole expansions in the upwards pass of the FMM algorithm. For this, the coefficients $L^\Delta_{\ell,\cellidxi}$ are initialised to zero for all cells $\cellidxi$ and levels $\ell$. Next, on each level we loop over all cells $\cellidxi$ and increment $\Psi^\Delta_{\ell,\cellidxj}$ for all $\cellidxj\in\operatorname{IL}(\cellidxi)$ by adding the contribution of all monopoles in $\cellidxi$ to the local expansion in $\cellidxj$. Since $N$ monopoles have to be considered on each level, the computational complexity of the setup phase is $\mathcal{O}(p^2LN) = \mathcal{O}(p^2N\log(N))$.
%%%%%%%%%%%%%%%%%%%%%%%%%%%%%%%%%%%%%%%%%%%%%%%%%%%%%%%%%%%%%%%%%%%
\subsection{Alternative algorithm based on multipole expansions}\label{sec:alternative_multipole_algorithm}
%%%%%%%%%%%%%%%%%%%%%%%%%%%%%%%%%%%%%%%%%%%%%%%%%%%%%%%%%%%%%%%%%%%
For reference, we now describe the alternative algorithm introduced in \cite{Hoeft2017}, which is based entirely on multipole expansions and which we also implemented for reference. In analogy to Equation \eqref{eqn:local_expansion}, we define the multipole expansion of all particles contained in box $\cellidxi$ on level $\ell$ around the centre of the box 
\begin{equation}
  \Phi_{\ell,\cellidxi}(\vec{\delta r}) = \sum_{n=0}^{p}\sum_{m=-n}^{+n} \left(M_{\ell,\cellidxi}\right)_n^m (\delta r)^{-(n+1)} Y_n^m(\delta \theta,\delta\phi)
  \qquad\text{with $(\delta r,\delta\theta,\delta\phi)=\textsf{spherical}(\vec{\delta r})$.}
  \label{eqn:multipole_expansion}
\end{equation}
Assuming that the particles in the box $\cellidxi$ have coordinates $\vec{\delta r}_i$ and charges $q_i$ with $i\in I_\cellidxi\subset \{1,\dots,N\}$, the explicit expression for the multipole expansion coefficients in Equation \eqref{eqn:multipole_expansion} is
\begin{equation}
  (M_{\ell,\cellidxi})_n^m = \sum_{i\in I_{\cellidxi}} q_i (\delta r_i)^{n}Y_n^{-m}(\delta\theta_i,\delta\phi_i)
  \qquad\text{with $(\delta r_i,\delta\theta_i,\delta\phi_i)=\textsf{spherical}(\vec{\delta r}_i)$.}
    \label{eqn:multipole_expansion_coefficients}
\end{equation}
Algorithms \ref{alg:propose_multipole} and \ref{alg:accept_multipole} describe how a potential move is proposed and accepted, using only multipole expansions; the two algorithms should be compared to Algorithms \ref{alg:propose_local} and \ref{alg:accept_local} above. Both methods have a computational complexity of $\mathcal{O}(p^2\log(N))$, but observe that the expensive loops over the interaction list are now carried out in the proposal step in Algorithm \ref{alg:propose_multipole}. Again, it would be possible to initialise the multipole expansion coefficients $M_{\ell,\cellidxi}$ at the beginning of the simulation in $\mathcal{O}(p^4N)$ time with one upward pass of the native FMM method. However, for simplicity we chose to calculate them directly by looping over the cells on all levels and accumulating the multipole coefficients from all particles in a particular cell using Equation \eqref{eqn:multipole_expansion_coefficients}; this is written down explicitly in Algorithm \ref{alg:initialise_multipole} which has $\mathcal{O}(p^2N\log(N))$ complexity. The resulting coefficients $(M_{\ell,\cellidxi})_n^m$ are not identical to those that would be have been obtained in the upward pass of the FMM algorithm, where they are calculated by recursively combining expansions on subsequent levels. However, the difference between the two ways of computing the multipole coefficients can be bounded as in the standard FMM error analysis.
\setcounter{algorithm}{0}
\renewcommand\thealgorithm{M.\arabic{algorithm}}
\begin{algorithm}
  \caption{Propose a move $\vec{r}\rightarrow\vec{r}'$ using multipole expansions. Input: initial position $\vec{r}$ of particle with charge $q$; target position $\vec{r}'$. Output: change in total electrostatic energy $\Delta U$}
  \label{alg:propose_multipole}
  \begin{algorithmic}[1]
    \STATE{Find fine-level cells $\cellidxi_L, \cellidxi'_L$ such that $\vec{r}\in\cellidxi_L$ and $\vec{r}'\in\cellidxi'_L$}
    \STATE{$\Delta U\mapsfrom 0$}
    \FOR{$\ell=L,L-1,\dots,1$}
    \FOR{$\cellidxj_\ell\in\text{IL}(\cellidxi_\ell)$ and $\cellidxj'_\ell\in\text{IL}(\cellidxi'_\ell)$}
    \STATE{Update $\Delta U\mapsfrom\Delta U + q\left(\Phi_{\ell,\cellidxj'_{\ell}}(\vec{r}'-\vec{R}_{\cellidxj'_\ell})- \Phi_{\ell,\cellidxj_{\ell}}(\vec{r}-\vec{R}_{\cellidxj_\ell})\right)$ using \textit{multipole} expansion in Equation \eqref{eqn:multipole_expansion}}
    \ENDFOR
    \IF{$\ell>1$}
    \STATE{Set $\cellidxi_{\ell-1}\mapsfrom \textsf{parent}(\cellidxi_\ell)$; $\cellidxi'_{\ell-1}\mapsfrom \textsf{parent}(\cellidxi'_\ell)$}
    \ENDIF
    \ENDFOR
    \STATE{Calculate change $\Delta U_{\text{direct}}$ in electrostatic energy from direct interactions with Algorithm \ref{alg:propose_direct}}
    \STATE{$\Delta U\mapsfrom\Delta U+\Delta U_{\text{direct}}$}
  \end{algorithmic}
\end{algorithm}
\begin{algorithm}
  \caption{Accept a move $\vec{r}\rightarrow\vec{r}'$ using multipole expansions. Input: old position $\vec{r}$, new position $\vec{r}'$}
  \label{alg:accept_multipole}
  \begin{algorithmic}[1]
    \STATE{Find fine-level cells $\cellidxi_L, \cellidxi'_L$ such that $\vec{r}\in\cellidxi_L$ and $\vec{r}'\in\cellidxi'_L$}
    \FOR{$\ell=L,L-1,\dots,1$}
    \STATE{Set $(\delta r,\delta\theta,\delta\phi) \mapsfrom \textsf{spherical}(\vec{r}-\vec{R}_{\cellidxi_\ell})$ and $(\delta r',\delta\theta',\delta\phi')\mapsfrom\textsf{spherical}(\vec{r}'-\vec{R}_{\cellidxi'_\ell})$}
    \FOR{$n=0,\dots,p$}
    \FOR{$m=-n,\dots,+n$}
    \STATE{Update $(M_{\ell,\cellidxi_\ell})_n^m \mapsfrom (M_{\ell,\cellidxi_\ell})_n^m - q (\delta r)^{n} Y_n^{-m}(\delta\theta,\delta\phi)$}
    \STATE{Update $(M_{\ell,\cellidxi'_\ell})_n^m \mapsfrom (M_{\ell,\cellidxi'_\ell})_n^m + q (\delta r')^{n} Y_n^{-m}(\delta\theta',\delta\phi')$}
    \ENDFOR
    \ENDFOR
    \IF{$\ell>1$}
    \STATE{Set $\cellidxi_{\ell-1}\mapsfrom \textsf{parent}(\cellidxi_\ell)$; $\cellidxi'_{\ell-1}\mapsfrom \textsf{parent}(\cellidxi'_\ell)$}
    \ENDIF
    \ENDFOR
  \end{algorithmic}
\end{algorithm}
\begin{algorithm}
  \caption{Initialise multipole expansion coefficients $M_n^m$ for electrostatic calculation with Algorithms \ref{alg:propose_multipole} and \ref{alg:accept_multipole}}
  \label{alg:initialise_multipole}
  \begin{algorithmic}[1]
  \FOR{levels $\ell=1,\dots,L$}
  \FORALL{cells $\cellidxi_\ell$ on level $\ell$}
  \FOR{$n=0,\dots,p$ and $m=-n\dots+n$}
  \STATE{Set $(M_{\ell,\cellidxi_\ell})_n^m = 0$}
  \ENDFOR
  \ENDFOR
  \FORALL{cells $\cellidxi_\ell$}
  \FORALL{particles with charge $q_i$ and position $\vec{r}_i\in\cellidxi_\ell$}
  \STATE{Set $(\delta r_i,\delta\theta_i,\delta\phi_i)\mapsfrom\textsf{spherical}(\vec{r}_i-\vec{R}_{\cellidxi_\ell})$}
  \FOR{$n=0,\dots,p$}
  \FOR{$m=-n,\dots,+n$}
  \STATE{Update $(M_{\ell,\cellidxi_\ell})_n^m \mapsfrom (M_{\ell,\cellidxi_\ell})_n^m + q_i (\delta r_i)^{n} Y_n^{-m}(\delta\theta_i,\delta\phi_i)$}
  \ENDFOR
  \ENDFOR
  \ENDFOR
  \ENDFOR
  \ENDFOR
  \end{algorithmic}
\end{algorithm}
\clearpage
%%%%%%%%%%%%%%%%%%%%%%%%%%%%%%%%%%%%%%%%%%%%%%%%%%%%%%%%%%%%%%%%%%%
\subsection{Boundary conditions}
%%%%%%%%%%%%%%%%%%%%%%%%%%%%%%%%%%%%%%%%%%%%%%%%%%%%%%%%%%%%%%%%%%%
So far we have implicitly assumed that free-space boundary conditions are used for the calculation of the electrostatic energy.
In this case the interaction lists on the coarsest two levels of Algorithms 
\ref{alg:accept_local}, \ref{alg:initialise_local}, \ref{alg:propose_multipole} and \ref{alg:initialise_local} are empty. This implies that $\Psi^{\Delta}_{1,1}=\Psi^{\Delta}_{2,\cellidxi}=\Phi_{1,1}=\Phi_{2,\cellidxi}=0$ and levels $\ell=1,2$ can be skipped when looping over the hierarchical tree. It is possible to adapt all algorithms in this section for simulations with periodic boundary conditions by making the following modifications:
\begin{itemize}
\item In Algorithms \ref{alg:accept_local} and \ref{alg:initialise_local}, extend the domain $\Omega$ by $3^3-1=26$ identical copies of the simulation cell to obtain an extended computational domain $\overline{\Omega}$. In the loop over $\cellidxi_\ell$ and $\cellidxi'_\ell$, include the copies of those cells in the extended domain $\overline{\Omega}$.
\item By following the approach described in detail in \cite[Section 3.1]{Saunders2019}, extend Algorithm \ref{alg:initialise_local} to initialise the data structures $K$ and $E$ required to compute the electrostatic contribution of all charges outside $\overline{\Omega}$.
  \begin{algorithmic}[1]
  \STATE{Set $K_n^m \mapsfrom 0,~E_n^m \mapsfrom 0~\forall~m,n$ }
  \FORALL{particles with charge $q_i$ and position $\vec{r}_i\in\cellidxi_\ell$}
  \STATE{Set $(\delta r_i,\delta\theta_i,\delta\phi_i)\mapsfrom\textsf{spherical}(\vec{r}_i-\vec{R}_{1})$}
  \FOR{$n=0,\dots,p$}
  \FOR{$m=-n,\dots,+n$}
  \STATE{Set $K_n^m \mapsfrom K_n^m + q_i(\delta r_i)^n Y_n^{-m}(\delta\theta_i,\delta\phi_i)$}
  \STATE{Set $E_n^m \mapsfrom E_n^m + q_i(\delta r_i)^n Y_n^{m}(\delta\theta_i,\delta\phi_i)$}
  \ENDFOR
  \ENDFOR
  \ENDFOR
  \STATE{Store $K$ and $E$.}
  \end{algorithmic}
\item Extend Algorithm \ref{alg:accept_local} to update $K$ and $E$ when a move is accepted.
    \begin{algorithmic}[1]
    \STATE{Set $(\delta r,\delta\theta,\delta\phi) \mapsfrom \textsf{spherical}(\vec{r}-\vec{R}_{1})$ and $(\delta r',\delta\theta',\delta\phi')\mapsfrom\textsf{spherical}(\vec{r}'-\vec{R}_{1})$}
    \FOR{$n=0,\dots,p$}
    \FOR{$m=-n,\dots,+n$}
    \STATE{Set $K_n^m \mapsfrom K_n^m + q\left((\delta r')^nY_n^{-m}(\delta\theta',\delta\phi')-(\delta r)^nY_n^{-m}(\delta\theta,\delta\phi)\right)$}
    \STATE{Set $E_n^m \mapsfrom E_n^m + q\left((\delta r')^nY_n^{m}(\delta\theta',\delta\phi')-(\delta r)^nY_n^{m}(\delta\theta,\delta\phi)\right)$}
    \ENDFOR
    \ENDFOR
  \end{algorithmic}
\item Extend Algorithm \ref{alg:propose_local} to include the contributions to the energy differences from the proposed move in periodic images outside $\overline{\Omega}$ by computing the proposed differences to $K$ and $E$ to determine the change in energy (the linear operator $\mathcal{R}$ is introduced in \cite[Section 2.2.1]{Saunders2019}).
    \begin{algorithmic}[1]
        \STATE{Set $H \mapsfrom \mathcal{R}(K)$}
        \STATE{Set $U^\infty \mapsfrom \sum_{n=0}^{p} \sum_{m=-n}^n E_n^m H_n^m$}
    \STATE{Set $(\delta r,\delta\theta,\delta\phi) \mapsfrom \textsf{spherical}(\vec{r}-\vec{R}_{1})$ and $(\delta r',\delta\theta',\delta\phi')\mapsfrom\textsf{spherical}(\vec{r}'-\vec{R}_{1})$}
    \FOR{$n=0,\dots,p$}
    \FOR{$m=-n,\dots,+n$}
    \STATE{Set $\delta K_n^m \mapsfrom K_n^m + q\left((\delta r')^nY_n^{-m}(\delta\theta',\delta\phi')-(\delta r)^nY_n^{-m}(\delta\theta,\delta\phi)\right)$}
    \STATE{Set $\delta E_n^m \mapsfrom E_n^m + q\left((\delta r')^nY_n^{m}(\delta\theta',\delta\phi')-(\delta r)^nY_n^{m}(\delta\theta,\delta\phi)\right)$}
    \ENDFOR
    \ENDFOR
    \STATE{Set $\delta H \mapsfrom \mathcal{R}(\delta K)$}
    \STATE{Set $\Delta U^\infty \mapsfrom \sum_{n=0}^{p} \sum_{m=-n}^n \delta E_n^m \delta H_n^m - U^\infty$}
    \STATE{Set $\Delta U \mapsfrom \Delta U + \Delta U ^\infty$}
  \end{algorithmic}
\item In Algorithm \ref{alg:propose_direct}, extend $\mathcal{N}_b(\alpha_L)$ and $\mathcal{N}_b(\alpha'_L)$ to include all cells in the extended domain $\overline{\Omega}$. Further, the self-interaction term has to be modified to take into account spurious interactions with the additional copies of the original particle. As discussed in detail in \cite[Section 3.1]{Saunders2019} this can be done by replacing the update in line 8 of Algorithm \ref{alg:propose_direct} by
  \begin{equation*}
    \Delta U_{\text{direct}} \mapsfrom \Delta U_{\text{direct}} - \sum_{\vec{\nu}\in [-1,0,+1]^3}\frac{q^2}{|\vec{r}' - (\vec{r}+a\vec{\nu})|}+\frac{q^2}{a}\left(6+\frac{8}{\sqrt{3}}+\frac{12}{\sqrt{2}}\right).
    \end{equation*}
\end{itemize}
Similar modifications have to be made to Algorithms \ref{alg:propose_multipole} and \ref{alg:initialise_multipole}. In practice, iteration over the 26 additional copies of the simulation cells can be implemented by modifying data structures such as neighbour lists, see \cite{Saunders2019} for a more detailed discussion.
%%%%%%%%%%%%%%%%%%%%%%%%%%%%%%%%%%%%%%%%%%%%%%%%%%%%%%%%%%%%%%%%%%%
\section{Implementation}\label{sec:implementation}
%%%%%%%%%%%%%%%%%%%%%%%%%%%%%%%%%%%%%%%%%%%%%%%%%%%%%%%%%%%%%%%%%%%
The algorithms described in this paper have been implemented as an extension to the performance portable framework for molecular dynamics (PPMD) described in \cite{Saunders2018}, which is freely available at\vspace{-1ex}\begin{center}\url{https://github.com/ppmd/ppmd}\end{center}\vspace{-1ex}
PPMD provides a high-level Python interface for particle-based simulations which require the efficient execution of user-defined operations over all particles or pairs of particles in a system. An obvious example of the latter is the calculation of inter-particle forces in molecular dynamics simulations, but the interface is sufficiently abstract to support more general operations such as the structure analysis algorithms discussed in Section 4 of \cite{Saunders2018}. PPMD automatically generates efficient code for executing short user-defined C-kernels for all particles or particle-pairs on different parallel architectures; both distributed- and shared- memory parallelism are supported and the code can run on non-standard architectures such as GPUs. Particle-specific data (such as e.g. charge, mass and velocity) is stored as instances of the Python \texttt{ParticleDat} class. Particle positions, which contain information that is relevant for parallel domain decompositions, are stored as instances of the specialised \texttt{PositionDat} class. In addition to electrostatic potential- and force- calculation with classical Ewald summation \cite{Ewald1921,Saunders2017a}, PPMD also contains an implementation of the standard FMM algorithm in three dimensions given in \cite{Greengard1997}. The PPMD framework therefore provides all necessary data structures for storing information (such as local- and multipole- expansion coefficients) on a nested hierarchy of grids which is required to implement the algorithms discussed in this paper. Algorithms \ref{alg:propose_local} - \ref{alg:initialise_local}, \ref{alg:propose_multipole} - \ref{alg:initialise_multipole} and \ref{alg:propose_direct} are implemented in the separate \verb+coulomb_mc+ Python package which is based on PPMD and can be downloaded from\vspace{-1ex}\begin{center}\url{https://github.com/ppmd/coulomb_mc}\end{center}\vspace{-1ex}
Algorithms \ref{alg:propose_local}, \ref{alg:accept_local}, \ref{alg:propose_multipole} and \ref{alg:accept_multipole} have been implemented as auto-generated C-code. This allows the pre-computation of constant expressions such as combinatorial factors that arise in the evaluation of the spherical harmonics and unrolling of nested loops such as the ones in lines 5-10 of Algorithm \ref{alg:accept_local} and lines 4-9 in Algorithm \ref{alg:accept_multipole}.  
Finally, the generated code is compiled for a specific chip architecture at runtime to ensure optimal performance. Currently our implementation supports cuboid geometries with free space- and periodic boundary conditions.
%%%%%%%%%%%%%%%%%%%%%%%%%%%%%%%%%%%%%%%%%%%%%%%%%%%%%%%%%%%%%%%%%%%
\subsection{FMM-MC user interface}
%%%%%%%%%%%%%%%%%%%%%%%%%%%%%%%%%%%%%%%%%%%%%%%%%%%%%%%%%%%%%%%%%%%
Recall that the local expansion coefficients $L_{\ell,\alpha}^\Delta$ which are required to compute (changes of) the electrostatic energy of the system of $N$ particles with charges $\{q_1,q_2,\dots,q_N\}$ and positions $\{\vec{r}_1,\vec{r}_2,\dots,\vec{r}_N\}$ are initialised with Algorithm \ref{alg:initialise_local}. In the \texttt{coulomb\_mc} package the charges of all particles are represented as a \texttt{PositionDat} instance \texttt{q}, whereas the positions are stored as a \texttt{ParticleDat} object \texttt{r}. At the beginning of the simulation the user populates \texttt{q} and \texttt{r} and uses those to create a \texttt{MCFMM\_LM} object, which is passed additional information on the domain (such as boundary conditions), the number of levels $L$  in the grid hierarchy and the number of expansion terms. The constructor of the \texttt{MCFMM\_LM} class then executes Algorithm \ref{alg:initialise_local} to initialise the values of the coefficients $L_{\ell,\alpha}^\Delta$ . Following this, Algorithm \ref{alg:propose_local} can be used to compute the change in electrostatic energy which occurs if particle $j$ transitions from its original position $\vec{r}=\vec{r}_j$ to a new position $\vec{r}'=\vec{r}'_j$. In the code this is realised by calling the \texttt{propose()} method of the \texttt{MCFMM\_LM} object and passing it the particle index $j$ and the new position $\vec{r}'=\vec{r}'_j$. Finally, Algorithm \ref{alg:accept_local} updates the local expansions $L_{\ell,\alpha}^\Delta$ if the proposed move $\vec{r}\mapsto\vec{r}'$ is accepted. Calling the class method \texttt{accept()}, which is passed the particle index $j$ and the new position $\vec{r}'_j$ of particle also executes Algorithm \ref{alg:accept_local} for the transition $\vec{r}_j\mapsto\vec{r}'_j$. It replaces the position of particle $j$ in the \texttt{PositionDat} \texttt{r} by $\vec{r}'_j$ and updates the expansion coefficients $L_{\ell,\alpha}^\Delta$

Listing \ref{lst:fmm-mc-interface} illustrates the creation of an \texttt{MCFMM\_LM} object, followed by the computation of the change in energy $\texttt{dU}=\Delta U$ which would result from moving particle $j=7$ to the new position $\texttt{rj\_new}=(0.1, 1.0, 5.2)$ by calling the \texttt{propose()} method. In the last line the move to the new position is accepted by calling the \texttt{accept()} method. Note that although the example in Listing \ref{lst:fmm-mc-interface} assumes that the proposed position is identical to the accepted position, this is not the case in general. Because of this and since the code keeps track of the \textit{total} energy of the system at each step, by default the \texttt{accept()} method executes Algorithm \ref{alg:propose_local} to compute the change in system energy $\Delta U$. This can be avoided by passing this change $\Delta U$ (which - as shown in Listing \ref{lst:fmm-mc-interface} - might have been computed in a previous call to \texttt{propose()} with the new position that is to be accepted) as an additional parameter to the \texttt{accept()} method.
\begin{figure}
\begin{center}
\begin{minipage}{1.0\linewidth}
  \begin{lstlisting}[language={[ppmd]{python}}, label=lst:fmm-mc-interface,caption={Illustration of how a FMM\_MC instance is created and then subsequently used to propose and accept moves.}]
# Create MCFMM_LM instance with positions in PositionDat r and charges in ParticleDat q.
# Here we use 3 tree levels and 12 expansion terms
MC = MCFMM_LM(r, q, domain, 'pbc', r=3, l=12)
# Perform the initial electrostatic solve
MC.initialise()

# Consider move of particle j=7 to new position rj_new = (0.1, 1.0, 5.2)
j = 7
rj_new = np.array((0.1, 1.0, 5.2))

# Propose a move of charge j to position rj_new
dU = MC.propose((j,rj_new))

# Accept a move of charge j to position rj_new
MC.accept((j, rj_new),dU)
\end{lstlisting}
\end{minipage}
\end{center}
\end{figure}
The corresponding multipole-based Algorithms \ref{alg:propose_multipole} - \ref{alg:initialise_multipole} can be used by creating an \texttt{MCFMM\_MM} object which keeps track of the multipole expansion coefficients $M_{\ell,\alpha}^\Delta$. The constructor of this class implements Algorithm \ref{alg:initialise_multipole}. The class methods \texttt{propose()} and \texttt{accept()} implement Algorithms \ref{alg:propose_multipole} and \ref{alg:accept_multipole} and can be used in exactly the same way as the corresponding methods of the \texttt{MCFMM\_MM} class described above.

Note that the aim of \verb+coulomb_mc+ is to provide functionality for the calculation of (changes in) the electrostatic energy through the high-level \texttt{MCFMM\_LM} and \texttt{MCFMM\_MM} classes, which typically dominates the runtime. It is up to the user to implement the overarching Monte Carlo algorithm which generates proposed new positions $\vec{r}'$ and uses the calculated energy differences to accept or reject particular moves, e.g. in a Metropolis Hastings step.
%%%%%%%%%%%%%%%%%%%%%%%%%%%%%%%%%%%%%%%%%%%%%%%%%%%%%%%%%%%%%%%%%%%
\section{Results}\label{sec:results}
%%%%%%%%%%%%%%%%%%%%%%%%%%%%%%%%%%%%%%%%%%%%%%%%%%%%%%%%%%%%%%%%%%%
In the following we quantify the performance of the algorithms introduced in Sections \ref{sec:MC_FMM_method} and \ref{sec:alternative_multipole_algorithm} and
implemented as described in Section \ref{sec:implementation}. We demonstrate numerically that, as expected, the time spent in each Monte Carlo step increases logarithmically with the number of particles in the system. To assess its overall performance, we also compare the runtime of our code to version 2.06 of the DL\_MONTE package \cite{Purton2013,Brukhno2019} which uses the classical Ewald method to compute electrostatic interactions.

All numerical experiments were carried out on the University of Bath \textit{``Balena''} HPC cluster. Compute nodes of this machine consist of two Intel E5-2650v2 CPUs, and all timing results are reported for sequential runs on a single core.
A snapshot of the source code which can be used to reproduce the results, along with all plotting scripts and raw data is provided at \cite{mcfmmpaper_zenodo}. The code was compiled using version 19.5.281 of the Intel compiler; DL\_MONTE was compiled with version 17.1.132 of the same compiler.
%%%%%%%%%%%%%%%%%%%%%%%%%%%%%%%%%%%%%%%%%%%%%%%%%%%%%%%%%%%%%%%%%%%
\subsection{Configuration and Parameter Selection} \label{sec:results_config}
%%%%%%%%%%%%%%%%%%%%%%%%%%%%%%%%%%%%%%%%%%%%%%%%%%%%%%%%%%%%%%%%%%%
The accuracy of the algorithms described in Sections \ref{sec:MC_FMM_method} and \ref{sec:alternative_multipole_algorithm} crucially depends on the number of local/multipole expansion terms, which can be quantified by $p$, the upper limit in the outer sum in Equations \eqref{eqn:local_expansion} and \eqref{eqn:multipole_expansion}. To provide a fair comparison between the methods introduced in this paper and the Ewald implementation in DL\_MONTE, $p$ is adjusted such that acceptance probabilities have errors which are comparable to those in DL\_MONTE.
For a proposed move $\vec{r} \rightarrow \vec{r}'$ which results in a change of energy of $\Delta U$, the relative error in the acceptance probability $\delta P$ is defined as
\begin{equation}
    \delta P = \frac{\left| P - P^*  \right|}{\left| P^* \right|}.\label{eqn:def_deltaP}
\end{equation}
Here $P=\exp\left( -\Delta U / (k_B T) \right)$ is the acceptance probability (computed by DL\_MONTE or an expansion based method) for a given choice of parameters.  Assuming that the exact change in energy is $\Delta U^*$, the exact acceptance probability is denoted as $P^*=\exp\left( -\Delta U^* / (k_B T) \right)$.
For a particular move $P^*$ is approximated to high accuracy by computing $\Delta U^*$ with the local expansion based method and $26$ expansion terms ($p=25$). 

The configuration for our numerical experiments is based on TEST01 \cite{Test01} in the DL\_POLY suite. This setup simulates a simple cubic NaCl crystal of alternating Sodium (Na) and Chloride (Cl) ions with a lattice constant of $a=$3.3\AA. Fully periodic boundary conditions are used for all numerical experiments, which are performed at a temperature of $T=273\mathrm{K}$.

To estimate the relative errors $\delta P$ in the acceptance probability, we start with an initial configuration of charges which is constructed by creating a cubic lattice of $22\times22\times22 = 10648$ ions as described in TEST01 and perturbing the initial position of each ion by adding a uniform random shift with a maximum size of $0.01a$ in each spatial direction. Based on this, 1000 moves are proposed (note that no moves are accepted) and for each move the acceptance probabilities $P$ for DL\_MONTE and the expansion based approaches are computed along with the ``exact'' acceptance probability $P^*$, which is estimated as described above. This process is repeated for 16 different initial configurations to generate a total of 16000 samples for the quantity $\delta P$ defined in Equation \eqref{eqn:def_deltaP}.% As a technical aside note that while for our implementation energy differences $\Delta U$ are readily available, computing $\Delta U$ is less straightforward for DL\_MONTE.

\begin{figure}[H]
    \includegraphics[scale=1.0]{\figdir/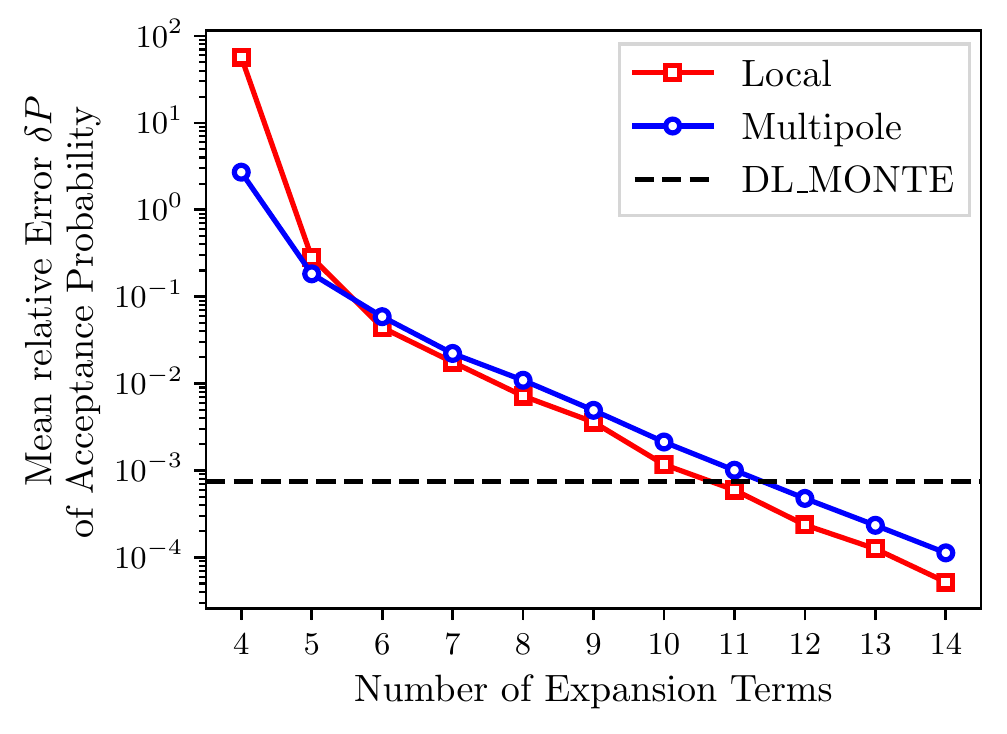}
    \centering
    \caption{Relative error $\delta P$ in acceptance probability for DL\_MONTE and the expansion based methods for a varying number of expansion terms.}
    \label{fig:dlm_acc_compare}
  \end{figure}

\begin{figure}[H]
    \includegraphics[scale=1.0]{\figdir/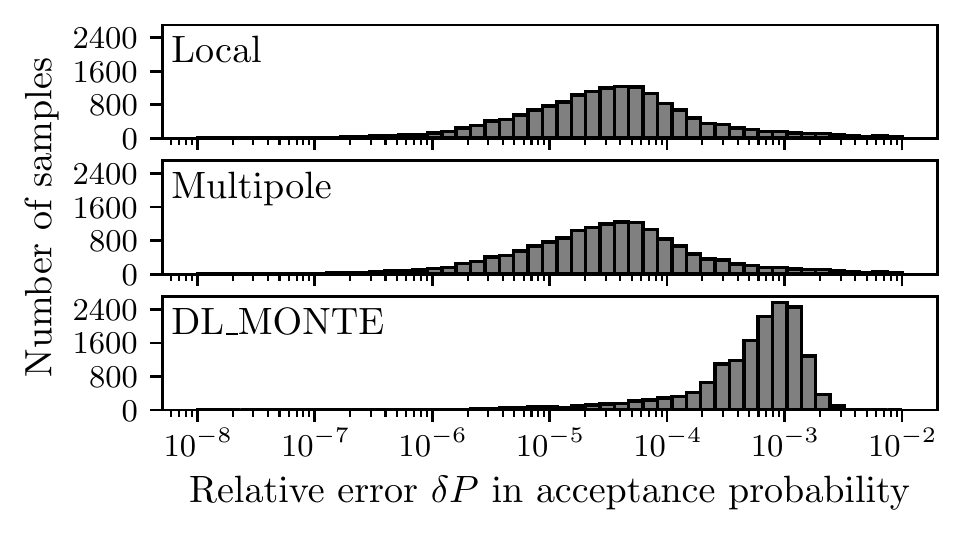}
    \centering
    \caption{Histograms of relative error $\delta P$ in acceptance probability. Results are shown for the local expansion based algorithm (top), the multipole expansion based algorithm (middle) and DL\_MONTE (bottom); 12 expansion terms ($p=11$) are used for first two methods.}
    \label{fig:dlm_prob_err_hists}
\end{figure}

\begin{figure}[H]
    \includegraphics[scale=1.0]{\figdir/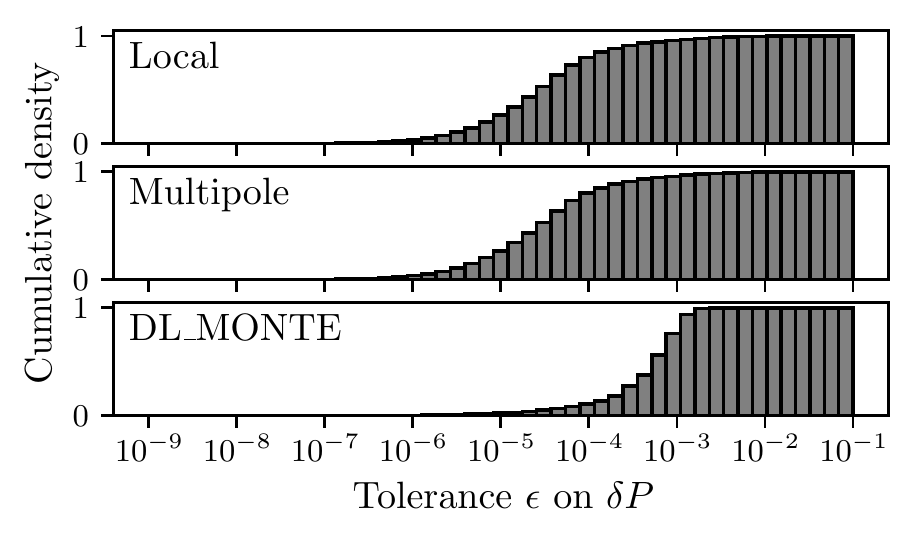}
    \centering
    \caption{Cumulative density of the relative error $\delta P$ in computed probabilities. Results are shown for the local expansion based algorithm (top), the multipole expansion based algorithm (middle) and DL\_MONTE (bottom); 12 expansion terms ($p=11$) are used for first two methods.}
    \label{fig:dlm_prob_cerr_hists}
\end{figure}

Figure \ref{fig:dlm_acc_compare} shows the mean relative error $\delta P$ (averaged over all 16000 samples) as a function of the number of expansion terms, which varies between 4 and 14 for the expansion based methods. This should be compared to the same quantity computed with DL\_MONTE at a fixed solver tolerance of $10^{-6}$, indicated by the horizontal dashed line. As those results show, choosing 12 expansion terms ($p=11$) results in a comparable mean relative error $\overline{\delta P}$ which is smaller than $10^{-3}$. Note that for a fixed value of $p$ the local expansion based method (Algorithm \ref{alg:propose_local}) has a slightly higher accuracy than the method which only uses multipole expansions (Algorithm \ref{alg:propose_multipole}).
  To compare the errors of all used methods in more detail, we also inspect the distribution of $\delta P$ over all 16000 samples. Figure \ref{fig:dlm_prob_err_hists} shows a histogram of the relative error in the acceptance probability, i.e. the number of samples which have a $\delta P$ that falls into a certain interval $[\delta P_1,\delta P_2]$. Results are shown both for DL\_MONTE (again using a solver tolerance of $10^{-6}$) and our expansion based methods with 12 expansion terms. The cumulative density of the probability distribution in Figure \ref{fig:dlm_prob_err_hists} is plotted in Figure \ref{fig:dlm_prob_cerr_hists}. In other words, for a given tolerance $\epsilon$ on $\delta P$, Figure \ref{fig:dlm_prob_cerr_hists} shows the percentage of samples that have a relative error which does not exceed $\epsilon$. As both figures demonstrate, the spread in errors in slightly larger for the expansion based methods: although for those methods a larger fraction of samples have errors well below the tolerance of $10^{-3}$, there is a small number of outliers. This, however, is consistent between the two expansion based methods. 
  
\begin{figure}[H]
    \includegraphics[scale=1.0]{\figdir/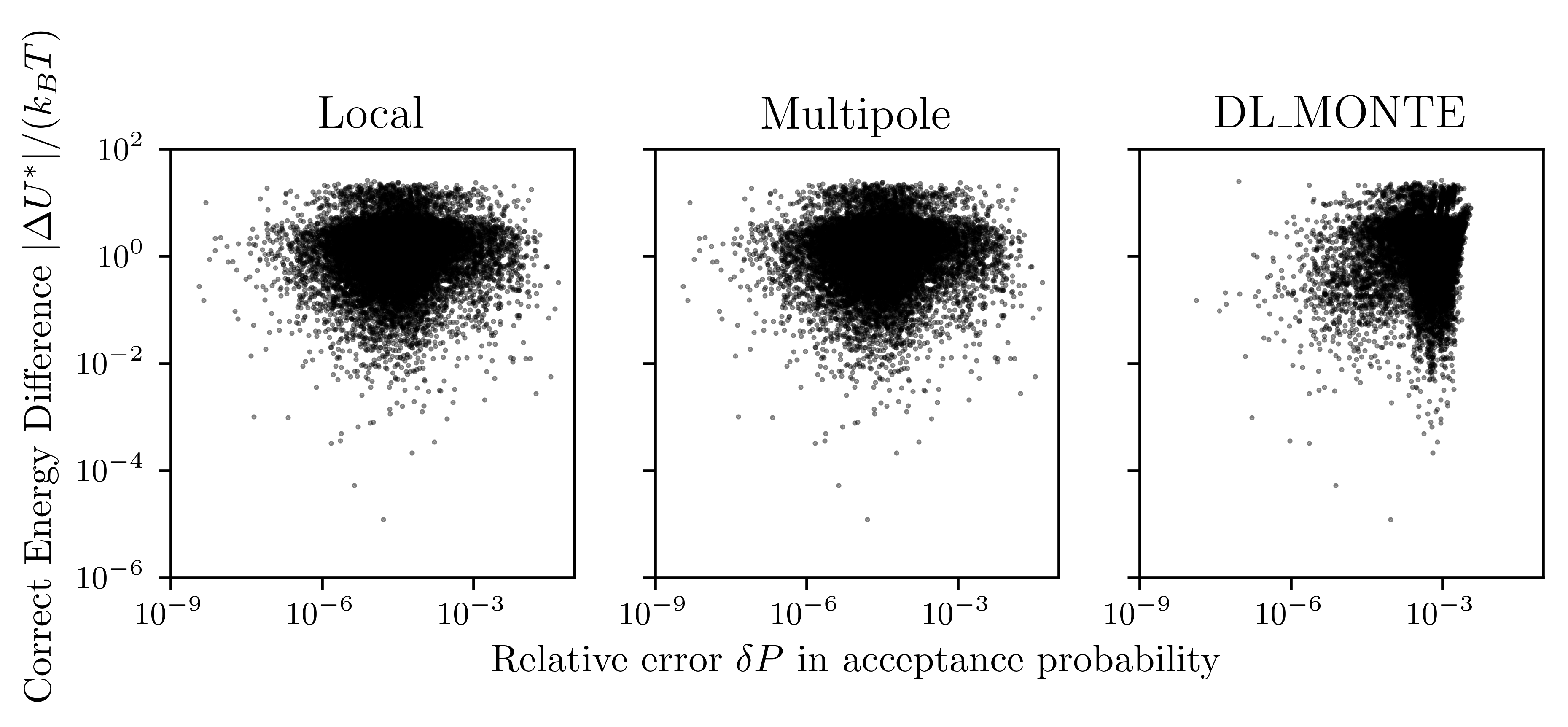}
    \centering
    \caption{Scatter plot of $\Delta U^*/(k_B T)$ against the relative error $\delta P$ in acceptance probability. Results are shown for the local expansion based algorithm (left), the multipole expansion based algorithm (middle) and DL\_MONTE (right); 12 expansion terms ($p=11$ ) are used for first two methods.}
    \label{fig:dlm_prob_err_scatter}
\end{figure}

Finally, observe that a large relative error $\delta P$ in the acceptance probability $P$ will only translate into a large absolute error on $P$ if $\Delta U^*$ is also large. It is therefore instructive to also produce a scatter plot of $\Delta U^*$ against $\delta P$ for all samples and this is shown in Figure \ref{fig:dlm_prob_err_scatter}.

%%%%%%%%%%%%%%%%%%%%%%%%%%%%%%%%%%%%%%%%%%%%%%%%%%%%%%%%%%%%%%%%%%%
\subsection{Computational complexity} \label{sec:results_complexity}
%%%%%%%%%%%%%%%%%%%%%%%%%%%%%%%%%%%%%%%%%%%%%%%%%%%%%%%%%%%%%%%%%%%
Next, we investigate the growth in computational cost as a function of the number of charges $N$. Formally the number of levels $L$ of the hierarchical tree is $\mathcal{O}(\log(N))$. The relative proportion of time spent evaluating the local and multipole expansions in the propose stage (Algorithms \ref{alg:propose_local} and \ref{alg:propose_multipole}), update of expansion coefficients (Algorithms \ref{alg:accept_local} and \ref{alg:accept_multipole}) and direct, nearest neighbour calculations (Algorithm \ref{alg:propose_direct}) can be controlled by setting $L=\lfloor\log_8(\alpha N)\rfloor$ and varying the constant $\alpha$ (here $\lfloor\cdot\rfloor$ denotes the floor function defined by $\lfloor x\rfloor=\max\{n\in\mathbb{N}: n\le x\}$). The optimal value of the parameter $\alpha$ depends on the computer hardware, the average acceptance rate $\nu$ and the number of expansion terms. We define the acceptance rate as
\begin{equation}
    \nu = \frac{\text{Number of accepted moves}}{\text{Number of proposed moves}}.
\end{equation}
Assuming that on average $\nu^{-1} = e\approx2.718$ proposals have to be generated for each accepted move, we find that for our setup and with 12 expansion terms ($p=11$), the best results are obtained for $\alpha=0.327$ when using the local expansion based method (Algorithms \ref{alg:propose_local}, \ref{alg:accept_local}) and $\alpha=0.138$ if the multipole method (Algorithms \ref{alg:propose_multipole}, \ref{alg:accept_multipole}) is used. This also implies that for a given value of $N$, the optimal number of levels for both methods differs by less than 0.5.

\begin{figure}
\begin{center}
    \includegraphics[scale=0.5]{\figdir/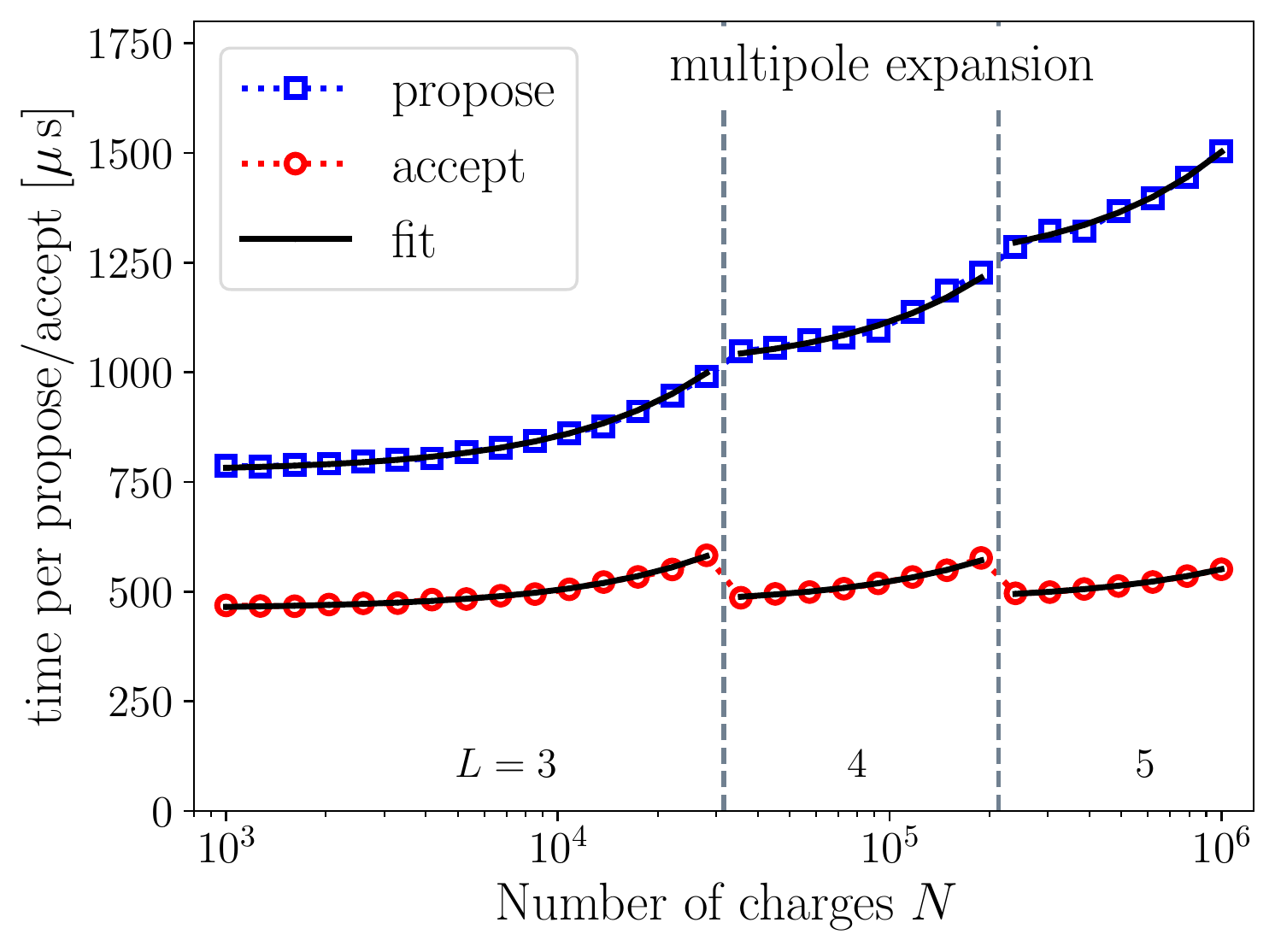}
    \includegraphics[scale=0.5]{\figdir/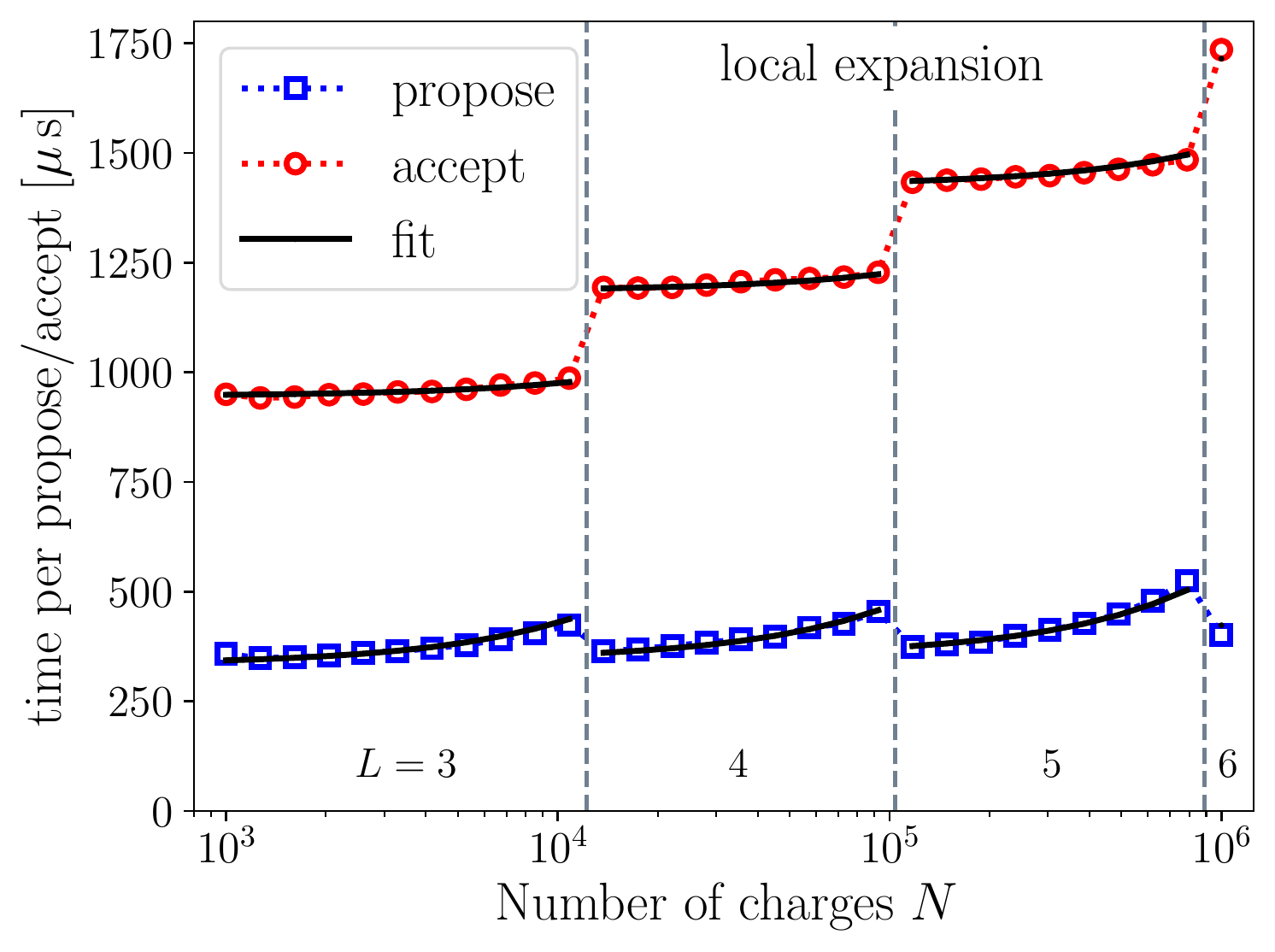}
    \caption{Time spent in the propose and accept operations as a function of the number of charges $N$ in the system. Results in the left figure were obtained with the multipole expansion based method in \cite{Hoeft2017} and show the time spent in Algorithm \ref{alg:propose_multipole} ($T_{\text{prop}}^{(\text{mp})}$, blue squares) and Algorithm \ref{alg:accept_multipole} ($T_{\text{acc}}^{(\text{mp})}$, red circles). The right figure shows the corresponding numbers for our new method based on local expansion; namely the time spent in Algorithm \ref{alg:propose_local} ($T_{\text{prop}}^{(\text{loc})}$, blue squares) and Algorithm \ref{alg:accept_local} ($T_{\text{acc}}^{(\text{loc})}$, red circles). The step-changes in measured times (marked by dashed vertical lines) correspond to increases in the number of levels $L$, which are shown at the bottom of each figure. The fit to the data (solid black lines) is discussed in Section \ref{results_complexity}.}
    \label{fig:scaling_with_N}
\end{center}
\end{figure}

To quantify the computational complexity of the propose stage (Algorithms \ref{alg:propose_local} and \ref{alg:propose_multipole}) and the accept stage (Algorithms \ref{alg:accept_local} and \ref{alg:accept_multipole}) separately, random proposals and accepted moves are generated for problems of increasing size, drawing particle positions and charges from a uniform random distribution with a maximal absolute displacement of $0.25\mathrm{\AA}$ in each spatial direction.
We investigated the computational cost of proposing and accepting moves for systems containing between a thousand ($N=10^3$) and a million ($N=10^6$) particles.
For each $N$ the initial arrangement of particles is constructed as described in Section \ref{sec:results_config}. Figure \ref{fig:scaling_with_N} shows the average time (measured over 1000 samples) per propose or accept operation as a function of the number $N$ of charges in the system; the fitted solid lines are discussed in Section \ref{results_complexity} below.
The measured times increase abruptly as the number of levels $L$ changes as incrementing the number of levels increases the number of expansions that must be evaluated and updated. Although asymptotically we expect all times to grow as $L\propto \log(N)$, there are significant differences in the rate of growth and absolute computational cost for the different implementations. While for the multipole based method from \cite{Hoeft2017} proposing a single move is significantly more expensive than accepting it, the opposite is true for our new method based on local expansions. The main reason for this is that the expensive loop over cells in the interaction list has to be executed in the propose stage on the of the multipole based method (Algorithm \ref{alg:propose_multipole}), whereas the interaction list is traversed in the accept stage of our new method (Algorithm \ref{alg:accept_local}). Overall we therefore expect our new method to be more efficient as the acceptance rate $\nu$ decreases, and the number of proposals is significantly larger than the number of accepted moves. In Metropolis Hastings simulations this is the case since the acceptance rate is usually significantly less than 1, a typical value is $1/e \approx 0.3679$.

Although one would naively expect the runtime of the multipole-based accept (Algorithm \ref{alg:accept_multipole}) and the local-based propose (Algorithm \ref{alg:propose_local}) to be roughly identical, the measured cost of the latter is slightly smaller.
Similarly, Figure 7 shows that the multipole-based propose (Algorithm \ref{alg:propose_multipole}) is slightly faster than the local-based accept (Algorithm \ref{alg:accept_local}).
This can be explained by details of the implementation.
Firstly, the bookkeeping operations for the propose- and accept stages introduce different overheads. For example, when proposing a move the cells $\alpha_\ell$, $\alpha_\ell'$ have to be computed and when a move is accepted data structures, such as the indirection map that assigns particles to cells, must be updated. Furthermore, in practice reading and writing floating point numbers carries a different cost. For example, in the loop over the interaction list in Algorithm \ref{alg:accept_local} expansion coefficients are updated, which requires a read and a write operation, however in Algorithm \ref{alg:propose_multipole} in the loop over the interaction list expansion coefficients are only read, which avoids any potential write contention. Finally bear in mind that the simulation uses periodic boundary conditions. In our implementation this introduces a small overhead when proposing a move and a slightly larger additional cost when a move is accepted.

To understand the growth of the runtime for increasing problem size $N$ in more detail observe the following: when an additional level is added to the octal tree as $N$ increases, initially there are $N/N_{\text{cell}} = \mathcal{O}(1)$ charges per cell on the finest level. However, as the total number of charges increases while the number of levels $L$ is kept fixed, the average number of charges per cell will grow. Hence the cost of Algorithms \ref{alg:propose_local} and \ref{alg:propose_multipole} grows in proportion to $N/N_{\text{cell}}$ for fixed $L$ since both require the computation of direct interactions with Algorithm \ref{alg:propose_direct}. In addition our implementation of Algorithms \ref{alg:accept_local} and \ref{alg:accept_local} contains a small number of bookkeeping operations which are formally $\mathcal{O}(N)$. As discussed in Section \ref{results_complexity},  the absolute contribution of those operations to the total runtime is very small (and will be amoritised by the cost of direct interactions induced by other short-range potentials such a repulsive Lennard-Jones field in real-life simulations). Ignoring this small $\mathcal{O}(N)$ contribution the asymptotic complexity of Algorithms \ref{alg:propose_local}, \ref{alg:accept_local}, \ref{alg:propose_multipole} and \ref{alg:accept_multipole} is $\mathcal{O}(L) = \mathcal{O}(\log(N))$.
%%%%%%%%%%%%%%%%%%%%%%%%%%%%%%%%%%%%%%%%%%%%%%%%%%%%%%%%%%%%%%%%%%%
\subsection{Comparison of full Monte Carlo simulation with DL\_MONTE} \label{sec:results_dlmonte}
%%%%%%%%%%%%%%%%%%%%%%%%%%%%%%%%%%%%%%%%%%%%%%%%%%%%%%%%%%%%%%%%%%%
Having quantified the time spent in the propose- and accept-stage of our expansion based algorithm separately, we now discuss the growth of the total runtime of an entire Monte Carlo simulation as a function of the problem size. For this we compare the performance of our expansion-based implementations in PPMD against DL\_MONTE.  Again we consider an NaCl crystal with the same initial arrangement of charges as described in Section \ref{sec:results_config}. To prevent oppositely charged ions from collapsing onto one another over the course of the simulation, a repulsive short-range Lennard-Jones potential with a fixed cutoff of $12\mathrm{\AA}$ is added. For the largest problem sizes (with $N\approx 10^5$ in a cubic box with a side length of $153.2\mathrm{\AA}$) this short-range potential adds an additional average cost of $0.18\mathrm{ms}$ per Monte Carlo step for our expansion based methods. This accounts for approximately $14\%$ of the total average cost per step for the local expansion implementation.

For each proposed move we create a random offset vector $\delta\vec{r}=\vec{r}'-\vec{r}$, such  that each component $\delta r_j$ is uniformly distributed in the interval $[-0.25\mathrm{\AA},+0.25\mathrm{\AA}]$. This resulted in an average acceptance rate of $\nu\approx 0.438$ for the expansion based methods. Note that the performance of the Ewald implementation in DL\_MONTE is not sensitive to the acceptance rate and that in DL\_MONTE the additional cost of accepting a proposed move is negligible. 

\begin{figure}[H]
\begin{center}
  \includegraphics[scale=0.9]{\figdir/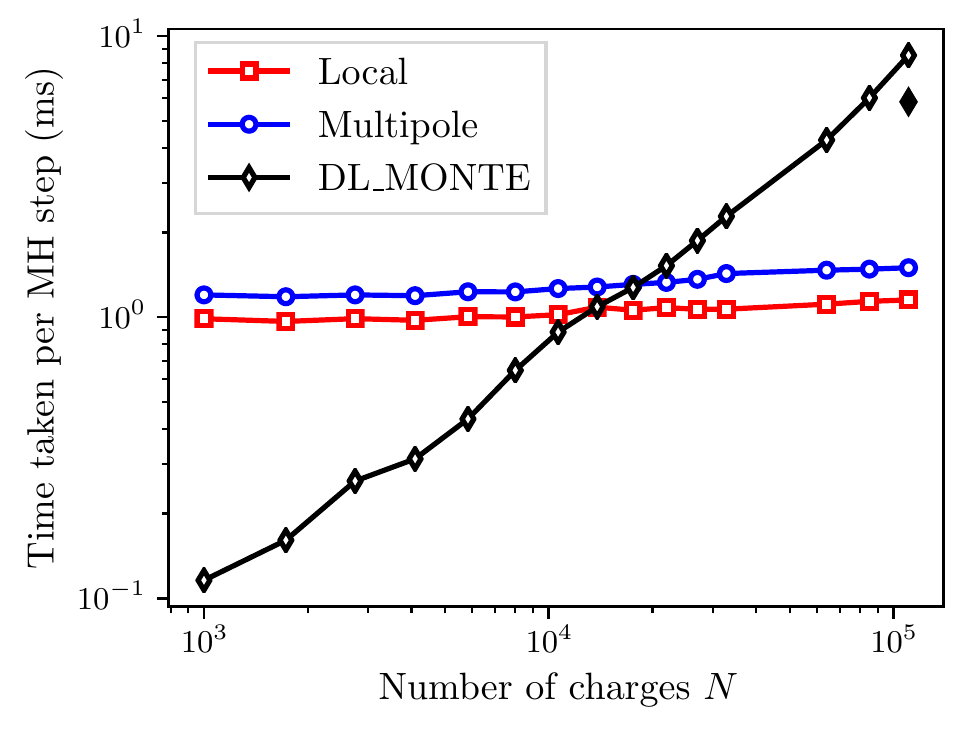}\hfill
  \includegraphics[scale=0.9]{\figdir/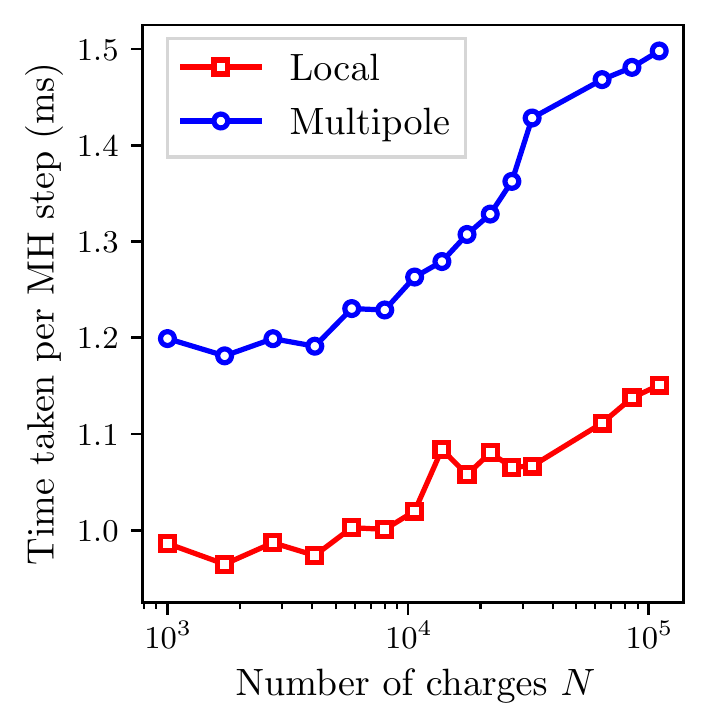}
    \caption{Time per Metropolis Hastings step for our implementations of the expansion based methods and DL\_MONTE. In the latter case the cutoff was kept fixed at $r_c=12\mathrm{\AA}$ for all problem sizes, except for the rightmost data point where results with an optimised cutoff of $r_c=28\mathrm{\AA}$ are also shown as a solid diamond. The plot on the right shows the same data but with a linear scale on the vertical axis.}
    \label{fig:dlm_comparison}
\end{center}
\end{figure}

Figure \ref{fig:dlm_comparison} shows the time per MC step for our implementations of the expansion based methods and for DL\_MONTE as a function of the number of charges $N$. The size of the system varies between $N=10^3$ and $N=10^5$ charges and the reported times are averaged over 1000 Metropolis-Hastings steps (i.e. 1000 proposed moves). We do not include the setup times, since those account for an insignificant fraction of the runtime for ``real'' simulations with a large number of Metropolis-Hastings steps.
For $N<10^4$ particles DL\_MONTE provides more performance than our implementations, but it is almost an order magnitude slower for the largest considered problem size ($N=10^5$).
Empirically the cost of the expansion based methods grows only relatively slowly with the problem size. For DL\_MONTE, on the other hand, the runtime is approximately proportional to the number of particles. This $\mathcal{O}(N)$ growth can be explained as follows: for the direct Ewald summation method with a short-range cutoff $r_c$, short-range interactions with $\mathcal{O}(\rho r_c^3)$ particles have to be computed for each proposed move in a system with density $\rho$. To balance errors in the short- and long-range computation, the physical cutoff in Fourier space and $r_c$ are inversely related. As a consequence, the number of Fourier modes that have to be considered grows like $\mathcal{O}(Vr_c^{-3})$. Hence, for fixed density where $V=\rho N$ the combined cost of the short- and long- range contributions is
\begin{equation}
  \mathrm{Cost}_{\mathrm{Ewald}}(r_c,N) = \mathcal{O}(r_c^3+Nr_c^{-3}).\label{eqn:cost_ewald}
\end{equation}
This cost can be minimised by varying the cutoff such that $r_c = r_c^{(0)}N^{1/6}$ for some constant $r_c^{(0)}$, as discussed for example in \cite{Kolafa1992} (see also \cite[Chapter~12]{Frenkel2001}), resulting in a total cost of $\mathrm{Cost}_{\mathrm{Ewald}}(r_c^{(0)}N^{1/6},N)=\mathcal{O}(N^{1/2})$ per proposal for the Ewald method. In the current version of DL\_MONTE the short-range cutoff for the Ewald summation has to be identical to the cutoff for Lennard-Jones interactions, which we fixed at $r_c=12\mathrm{\AA}$ to obtain the majority of the results in Figure \ref{fig:dlm_comparison}. In addition we also quantify how the results would change if the optimal short range cutoff is chosen. At the moment this requires manual fine-tuning and for practical reasons it was not possible to do this for all problem sizes. Equation \eqref{eqn:cost_ewald} shows that for a fixed cutoff of $r_c=12\mathrm{\AA}$ the cost of the Ewald summation is dominated by the long range contribution and grows in proportion to $N$. While this is a current limitation of DL\_MONTE and not a fundamental issue, it is worth stressing that even if this problem is overcome, Ewald summation has a computational complexity of $\mathcal{O}(N^{1/2})$ compared to the $\mathcal{O}(\log(N))$ complexity of our hierarchical methods. To demonstrate the effect of a more optimal short-range cutoff we varied the cutoff radius $r_c$ between $12\mathrm{\AA}$ and $32\mathrm{\AA}$ and found that a cutoff of $r_c = 28\mathrm{\AA}$ gives near optimal results. As shown by the rightmost datapoints in Figure \ref{fig:dlm_comparison}, where the results obtained with $r_c=28\mathrm{\AA}$ are indicated by a solid diamond, 
this reduces the time per MH step from $8.0\mathrm{ms}$ to $5.8\mathrm{ms}$ for $N\approx 10^5$ charges. This value might be reduced further in future releases of DL\_MONTE, if the different cutoffs can be varied independently to avoid the evaluation of short range potentials with an unnaturally large cutoff. With an optimally tuned cutoff the Ewald crossover between the Ewald- based method and our hierarchical algorithms would occur for larger problem sizes.

Finally, note that in its current implementation the setup cost of DL\_MONTE grows like $\mathcal{O}(N^2)$ instead of $\mathcal{O}(N^{3/2})$ which could be achieved for an optimal $r_c\propto N^{1/6}$. Although not considered here, this $\mathcal{O}(N^2)$ setup time can become computationally significant for systems containing a large number of charges. 
%%%%%%%%%%%%%%%%%%%%%%%%%%%%%%%%%%%%%%%%%%%%%%%%%%%%%%%%%%%%%%%%%%%
\subsection{Empirical results for growth of runtime} \label{results_complexity}
%%%%%%%%%%%%%%%%%%%%%%%%%%%%%%%%%%%%%%%%%%%%%%%%%%%%%%%%%%%%%%%%%%%
As will be shown below, the average speedup of our method relative to the one in \cite{Hoeft2017} is $1.25$ over all considered problem sizes. The smallest measured speedup is $1.18$, while the maximal speedup is $1.34$, with a median value of $1.23$. To model the speedup for large $N$ and establish an upper limit on the performance gains as $N\rightarrow\infty$ we fit the measured times in Figure \ref{fig:scaling_with_N} to
\begin{equation}
T(N) = \tau + \gamma L + \kappa \frac{N}{N_{\text{cell}}} + \sigma N
  \label{eqn:performance_model}
\end{equation}
with a least squares approach. In this expression $L$ is the number of levels of the octree hierarchy for a given value of $N$ and $N_{\text{cell}}=8^{L-1}$ is the number of cells on the finest level for this octree depth. The numerical values of the four fit parameters $\tau$, $\gamma$, $\kappa$ and $\sigma$ for the propose- and accept stages of the multipole- and local expansion based methods are given in Table \ref{eqn:fit_coefficients}. As the solid curves in Figure \ref{fig:scaling_with_N} show, Equation \eqref{eqn:performance_model} models the data extremely well. As expected the coefficient of the $\mathcal{O}(N)$ term is very small.
\begin{table}
  \begin{center}
\begin{tabular}{llrrrr}
\hline expansion & stage & \multicolumn{1}{c}{$\tau$} & \multicolumn{1}{c}{$\gamma$} & \multicolumn{1}{c}{$\kappa$} & \multicolumn{1}{c}{$\sigma$}\\\hline\hline
     \multirow{2}{*}{multipole}  &   propose & $  89.15$ & $ 228.40$ & $   0.50$ & $ 1.49\cdot10^{-4}$ \\ 
                                 &    accept & $ 436.71$ & $   8.12$ & $   0.27$ & $ 7.11\cdot10^{-6}$ \\ 
\hline
         \multirow{2}{*}{local}  &   propose & $ 304.01$ & $   9.92$ & $   0.61$ & $ 4.12\cdot10^{-5}$ \\ 
                                 &    accept & $ 225.80$ & $ 239.99$ & $   0.19$ & $ 4.37\cdot10^{-5}$ \\ 
\hline
\end{tabular}
    \caption{Numerical values of fit parameters for empirical performance model in Equation \eqref{eqn:performance_model}.}\label{eqn:fit_coefficients}
    \end{center}
  \end{table}
The empirical model in Eq. \eqref{eqn:performance_model} allows to predict the expected speedup for a given problem size $N$ and acceptance rate $\nu$ as
\begin{equation}
S(N) = \frac{T_{\text{prop}}^{(\text{mp})}(N)+\nu T_{\text{acc}}^{(\text{mp})}(N)}{T_{\text{prop}}^{(\text{loc})}(N)+\nu T_{\text{acc}}^{(\text{loc})}(N)}\label{eqn:speedup}
\end{equation}
where superscripts ``(loc)'' and ``(mp)'' denote the local and multipole expansion based methods and subscripts ``prop'' and ``acc'' label propose and accept operations respectively.
Ignoring the $\mathcal{O}(N)$ term in Equation \eqref{eqn:performance_model} (this can be justified by the small value of $\sigma$ in Table \ref{eqn:fit_coefficients}), the asymptotic speedup is
\begin{equation}
S_\infty = \lim_{N\rightarrow \infty}S(N) = \frac{ \gamma_{\text{prop}}^{(\text{mp})}+\nu\gamma_{\text{acc}}^{(\text{mp})}}{\gamma_{\text{prop}}^{(\text{loc})}+\nu \gamma_{\text{acc}}^{(\text{loc})}}\label{eqn:speedup_asymptotic}
\end{equation}

Further observe that asymptotically the relative cost of Algorithms \ref{alg:propose_local} and \ref{alg:propose_multipole} is $\gamma_{\text{prop}}^{(\text{mp})}/\gamma_{\text{prop}}^{(\text{loc})} = 23.03$ while the relative time spent in Algorithms \ref{alg:accept_multipole} and \ref{alg:accept_local} is $\gamma_{\text{acc}}^{(\text{loc})}/\gamma_{\text{acc}}^{(\text{mp})} = 29.56$. These ratios deviate from the theoretically expected value of $189$ (which, as the reader will recall, is the number of cells in the interaction list) due to details of the implementation. In particular looping over a larger set of cells in the interaction lists allows the compiler to carry out additional optimisation and vectorise the code.

\begin{figure}
\begin{center}
    \includegraphics[scale=0.65]{\figdir/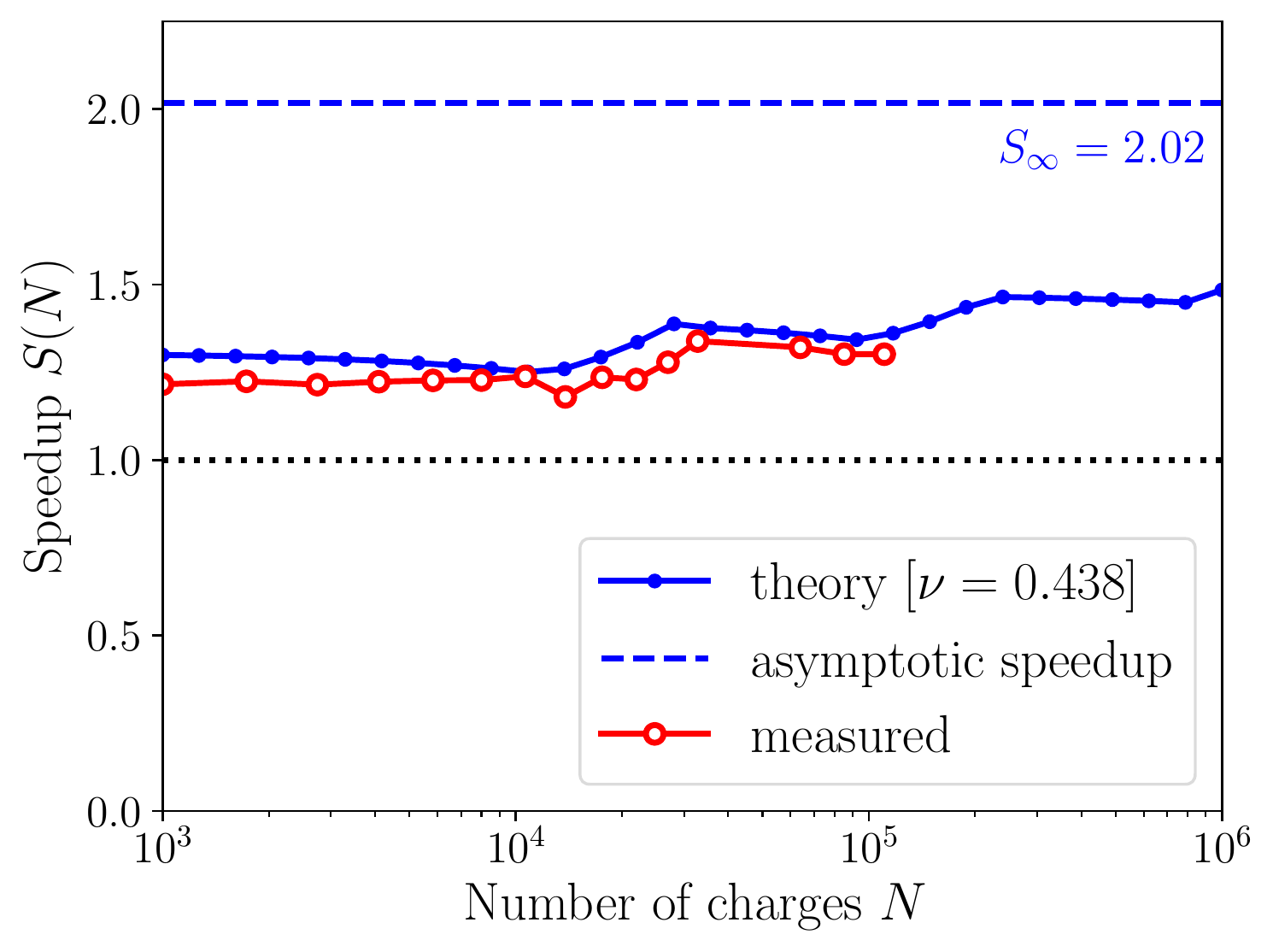}
    \caption{Theoretical speedup $S(N)$ (solid blue line) as defined in Equations \eqref{eqn:performance_model} and \eqref{eqn:speedup} for $\nu=0.438$; the asymptotic speedup $S_\infty$ (dash blue line), is given in Equation \eqref{eqn:speedup_asymptotic}. The measured speedup, in solid red, is computed from the local and multipole timings plotted in Figure \ref{fig:dlm_comparison}.}
    \label{fig:speedup}
\end{center}
\end{figure}

In Figure \ref{fig:speedup} we plot the theoretical and measured speedup for an acceptance rate $\nu = 0.438$. Here the theoretical speedups (blue) predicted with Equation \eqref{eqn:performance_model} are computed from the time per propose and accept operation computed in Section \ref{sec:results_complexity} and fitted with the model in Equation \eqref{eqn:performance_model}. The measured speedups (red) are obtained from the data visualised in Figure \ref{fig:dlm_comparison}. The theoretical upper limit $S_\infty$ for the speedup defined in Equation \eqref{eqn:speedup_asymptotic} is also shown; for the fitted values of $\gamma$ given in Table \ref{eqn:fit_coefficients} and with $\nu = 0.438$, we find $S_\infty = 2.02$.
As Figure \ref{fig:speedup} demonstrates, the fit reproduces the measured speedup, which lies in the range $[1.18,1.34]$, reasonably well. The expected asymptotic speedup for large $N$ is around 2, demonstrating that our method has the potential to approximately halve the runtime compared to the algorithm in \cite{Hoeft2017}. However, this speedup is only reached for values of $N$ which are significantly larger than the ones considered in this work; for $N=10^6$ we expect a speedup of roughtly $1.5\times$.
%%%%%%%%%%%%%%%%%%%%%%%%%%%%%%%%%%%%%%%%%%%%%%%%%%%%%%%%%%%%%%%%%%%
\section{Conclusion}\label{sec:conclusion}
%%%%%%%%%%%%%%%%%%%%%%%%%%%%%%%%%%%%%%%%%%%%%%%%%%%%%%%%%%%%%%%%%%%
In this paper we presented a new hierarchical method for accurately including electrostatic interactions in Monte Carlo simulations. Our algorithm has a computational complexity of $\mathcal{O}(\log(N))$ per Metropolis Hastings step. Compared to the related technique in \cite{Hoeft2017}, our method reduces the average cost of a MH step as the balance of work between the propose and accept operations is more favourable for typical acceptance rates. Numerically we find that runtimes are reduced by around $30\%$ for systems with $N=10^5$ charges, with the potential of a speedup of around $2\times$ for larger values of $N$. We demonstrated numerically that our implementation will effectively scale to systems containing $10^{6}$ charges whilst maintaining the expected computational complexity of $\mathcal{O}(\log(N))$ per MH step.
As the direct Ewald summation technique has a higher complexity of at least $\mathcal{O}(\sqrt{N})$, our implementation becomes more efficient for simulations with more than $10^4$ particles: for $N=10^5$ it is about an order of magnitude faster than the current DL\_MONTE implementation.

There are several avenues for future work. One obvious shortcoming of the present implementation is the lack of parallelisation. While Monte Carlo simulations are ``embarrassingly parallel'', and several Markov Chains can be generated in parallel without communications, this increases memory requirements. In our method this issue could be reduced to a certain degree by parallelising over the levels in the grid hierarchy. This is possible since the cost on each level is constant, and computations can be carried out independently, before summing the total energy in the propose stage.

Furthermore, here we have only considered single-particle moves and future work should also investigate other Monte Carlo transitions. For example, in a canonical ensemble the pressure is fixed but the volume of the simulation cell varies. In this case a proposed move could consist of a change of the system volume. In the worst case scenario, the energy change of such a proposed volume move is computed with a full solve of the electrostatic energy of the proposed state. When using an Ewald based approach this system solve incurs an $\mathcal{O}(N^\frac{3}{2})$ cost per MH step, however, with multipole- or local expansion based approaches this can potentially be reduced to an $\mathcal{O}(N)$ cost.
%%%%%%%%%%%%%%%%%%%%%%%%%%%%%%%%%%%%%%%%%%%%%%%%%%%%%%%%%%%%%%%%%%%
\section*{Acknowledgements}
%%%%%%%%%%%%%%%%%%%%%%%%%%%%%%%%%%%%%%%%%%%%%%%%%%%%%%%%%%%%%%%%%%%
This research made use of the Balena High Performance Computing (HPC) Service at the University of Bath. This project has received funding from the European Union’s Horizon 2020 research and innovation programme under grant agreements No 646176 and No 824158.
\FloatBarrier
\appendix
\ifbool{PREPRINT}{ % PREPRINT
%\end{multicols}
}{} % PREPRINT
\bibliographystyle{elsarticle-num}

\end{document}